\documentclass[final,5p,times,twocolumn,authoryear]{elsarticle}

\usepackage{amssymb}
\usepackage{lineno}
\usepackage{doi}
\usepackage{xspace}
\usepackage{natbib}
\usepackage{enumerate}
\newcommand\pnas{{Proc. Natl. Acad. Sci. }}
\newcommand\aj{{Astron. J. }}%
\newcommand\araa{{Annu. Rev. Astron. Astrophys. }}%
\newcommand\apj{{Astrophys. J. }}%
\newcommand\apjl{{Astrophys. J. Lett. }}%
\newcommand\aap{{Astron. Astrophys. }}%
\newcommand\icarus{{Icarus }}%
\newcommand\mnras{{Mon. Not. R. Astron. Soc. }}%
\newcommand\nat{{Nature }}%
\newcommand\gca{{Geochim.~Cosmochim.~Acta }}%
\newcommand\jgr{{J.~Geophys.~Res. }}%
\newcommand{\al}{$^{26}$Al\xspace}

\newcommand{\alr}{($^{26}$Al/$^{27}$Al)$_{\mathrm{form}}$\xspace}

\newcommand{\tform}{$t_{\mathrm{form}}$\xspace}
\newcommand{\tmin}{$t_{\mathrm{min}}$\xspace}
\newcommand{\tcollision}{$t_{\mathrm{collision}}$\xspace}
\newcommand{\tmax}{$t_{\mathrm{max}}$\xspace}
\newcommand{\RP}{$R_{\mathrm{P}}$\xspace}
\newcommand{\Rmax}{$R_{\mathrm{max}}$\xspace}

\newcommand{\Tsol}{$T_{\mathrm{sol}}$\xspace}

\newcommand{\Tcenter}{$T_{\mathrm{center}}$\xspace}
\newcommand{\dmath}{\mathrm{d}}
\newcommand{\RPmath}{R_{\mathrm{P}}}
\newcommand{\Tchondrule}{$T_{\mathrm{chondrule}}$\xspace}
\newcommand{\Tpost}{$T_{\mathrm{post}}$\xspace}
\newcommand{\phicrit}{$\varphi_{\mathrm{crit}}$\xspace}
\newcommand{\rmu}[1]{_{\mathrm{#1}}}

\journal{Icarus}

\begin{document}

\begin{frontmatter}

%% Title, authors and addresses

%% use the tnoteref command within \title for footnotes;
%% use the tnotetext command for theassociated footnote;
%% use the fnref command within \author or \address for footnotes;
%% use the fntext command for theassociated footnote;
%% use the corref command within \author for corresponding author footnotes;
%% use the cortext command for theassociated footnote;
%% use the ead command for the email address,
%% and the form \ead[url] for the home page:
%% \title{Title\tnoteref{label1}}
%% \tnotetext[label1]{}
%% \author{Name\corref{cor1}\fnref{label2}}
%% \ead{email address}
%% \ead[url]{home page}
%% \fntext[label2]{}
%% \cortext[cor1]{}
%% \address{Address\fnref{label3}}
%% \fntext[label3]{}

% \title{A thermomechanical `Goldilocks' regime for and from impact splash chondrule formation}
\title{Impact splash chondrule formation during planetesimal recycling}

\author[inst1,inst2]{Tim Lichtenberg\corref{cor1}}
\author[inst3]{Gregor J. Golabek}
\author[inst4]{Cornelis P. Dullemond}
\author[inst5]{Maria Sch{\"o}nb{\"a}chler}
\author[inst1]{Taras V. Gerya}
\author[inst2,inst6]{Michael R. Meyer}

\cortext[cor1]{Corresponding author. E-mail: \href{mailto:tim.lichtenberg@phys.ethz.ch}{tim.lichtenberg@phys.ethz.ch}.}

\address[inst1]{Institute of Geophysics, ETH Z{\"u}rich, Sonneggstrasse 5, 8092 Z{\"u}rich, Switzerland}
\address[inst2]{Institute for Astronomy, ETH Z{\"u}rich, Wolfgang-Pauli-Strasse 27, 8093 Z{\"u}rich, Switzerland}
\address[inst3]{Bayerisches Geoinstitut, University of Bayreuth, Universit{\"a}tsstrasse 30, 95440 Bayreuth, Germany}
\address[inst4]{Institute for Theoretical Astrophysics, Zentrum f{\"u}r Astronomie, Heidelberg University, Albert-Ueberle-Strasse 2, 69120 Heidelberg, Germany}
\address[inst5]{Institute of Geochemistry and Petrology, ETH Z{\"u}rich, Clausiusstrasse 25, 8092 Z{\"u}rich, Switzerland}
\address[inst6]{Department of Astronomy, University of Michigan, 1085 S. University Avenue, Ann Arbor, MI 48109, USA}

\begin{abstract}
Chondrules, mm-sized igneous-textured spherules, are the dominant bulk silicate constituent of chondritic meteorites and originate from highly energetic, local processes during the first million years after the birth of the Sun. So far, an astrophysically consistent chondrule formation scenario explaining major chemical, isotopic and textural features, in particular Fe,Ni metal abundances, bulk Fe/Mg ratios and intra-chondrite chemical and isotopic diversity, remains elusive. Here, we examine the prospect of forming chondrules from impact splashes among planetesimals heated by radioactive decay of short-lived radionuclides using thermomechanical models of their interior evolution. We show that intensely melted planetesimals with interior magma oceans became rapidly chemically equilibrated and physically differentiated. Therefore, collisional interactions among such bodies would have resulted in chondrule-like but basaltic spherules, which are not observed in the meteoritic record. This inconsistency with the expected dynamical interactions hints at an incomplete understanding of the planetary growth regime during the lifetime of the solar protoplanetary disk. To resolve this conundrum, we examine how the observed chemical and isotopic features of chondrules constrain the dynamical environment of accreting chondrite parent bodies by interpreting the meteoritic record as an impact-generated proxy of early solar system planetesimals that underwent repeated collision and reaccretion cycles. Using a coupled evolution-collision model we demonstrate that the vast majority of collisional debris feeding the asteroid main belt must be derived from planetesimals which were partially molten at maximum. Therefore, the precursors of chondrite parent bodies either formed primarily small, from sub-canonical aluminum-26 reservoirs, or collisional destruction mechanisms were efficient enough to shatter planetesimals before they reached the magma ocean phase. Finally, we outline the window in parameter space for which chondrule formation from planetesimal collisions can be reconciled with the meteoritic record and how our results can be used to further constrain early solar system dynamics.
\end{abstract}


\end{frontmatter}

%% \linenumbers

%%%%%%%%%%%%%%%%%%%%%%%%%%%%%%%%%%%%%%%%%%%%%%%%%%%%%%%%%%%%%%%%%%%%%%%%%%%%%%%%%%%%%%%%%%%%%%
%%%%%%%%%%%%%%%%%%%%%%%%%%%%%%%%%%%%%%%%%%%%%%%%%%%%%%%%%%%%%%%%%%%%%%%%%%%%%%%%%%%%%%%%%%%%%%
%%%%%%%%%%%%%%%%%%%%%%%%%%%%%%%%%%%%%%%%%%%%%%%%%%%%%%%%%%%%%%%%%%%%%%%%%%%%%%%%%%%%%%%%%%%%%%
%%%%%%%%%%%%%%%%%%%%%%%%%%%%%%%%%%%%%%%%%%%%%%%%%%%%%%%%%%%%%%%%%%%%%%%%%%%%%%%%%%%%%%%%%%%%%%
%%%%%%%%%%%%%%%%%%%%%%%%%%%%%%%%%%%%%%%%%%%%%%%%%%%%%%%%%%%%%%%%%%%%%%%%%%%%%%%%%%%%%%%%%%%%%%
%%%%%%%%%%%%%%%%%%%%%%%%%%%%%%%%%%%%% INTRODUCTION %%%%%%%%%%%%%%%%%%%%%%%%%%%%%%%%%%%%%%%%%%%
%%%%%%%%%%%%%%%%%%%%%%%%%%%%%%%%%%%%%%%%%%%%%%%%%%%%%%%%%%%%%%%%%%%%%%%%%%%%%%%%%%%%%%%%%%%%%%
%%%%%%%%%%%%%%%%%%%%%%%%%%%%%%%%%%%%%%%%%%%%%%%%%%%%%%%%%%%%%%%%%%%%%%%%%%%%%%%%%%%%%%%%%%%%%%
%%%%%%%%%%%%%%%%%%%%%%%%%%%%%%%%%%%%%%%%%%%%%%%%%%%%%%%%%%%%%%%%%%%%%%%%%%%%%%%%%%%%%%%%%%%%%%
%%%%%%%%%%%%%%%%%%%%%%%%%%%%%%%%%%%%%%%%%%%%%%%%%%%%%%%%%%%%%%%%%%%%%%%%%%%%%%%%%%%%%%%%%%%%%%
%%%%%%%%%%%%%%%%%%%%%%%%%%%%%%%%%%%%%%%%%%%%%%%%%%%%%%%%%%%%%%%%%%%%%%%%%%%%%%%%%%%%%%%%%%%%%%
\section{Introduction} 
\label{sec:intro}

Chondrules are igneous-textured spherules, typically 0.1--2 mm in diameter, and largely composed of the silicate minerals olivine and pyroxene. They are abundantly found in chondritic meteorites, together with other disk materials, such as Ca,Al-rich inclusions (CAIs) and the fine-grained matrix that includes presolar grains and primitive organics \citep{2014mcp..book...65S}. Chondrules are often surrounded by or close to beads of Fe,Ni metal \citep[e.g.,][]{2010GeCoA..74.2212W,2012M&PS...47.1176J} and show specific features, such as high abundances of moderately volatile elements like Na, K and S \citep{2008Sci...320.1617A,2014mcp..book...65S,2016JGRE..121.1885C} and diverse chemical and isotopic signatures \citep{2009GeCoA..73.5854J,2010E&PSL.294...85H,olsen2016magnesium}. Their peak temperatures were $\sim$ 1900 K or higher \citep{2008Sci...320.1617A,2016JGRE..121.1885C} with subsequent cooling in minutes to days \citep[e.g.,][]{2012GeCoA781H,2012M&PS...47.1139D,2012GeCoA98140W}. Most chondrules were formed during the earliest phases of the solar system within the first 3--4 million years after the formation of CAIs \citep[e.g.,][]{2009Sci...325..985V,2012Sci...338..651C} and show clear evidence for multiple melting cycles \citep[][and references therein]{2017LPICo1963.2006R}.

Because of their enigmatic features coupled with high-energy processing, chondrule formation is considered to be intimately linked to the physical processes in the protoplanetary disk or planetary accretion and spawned a multitude of proposed formation mechanisms. The often underlying view of how chondrules are intertwined with the planet formation process is that they were formed before accretion and therefore represent the fundamental building materials of the planets and asteroids \citep{2016JGRE..121.1885C}. In this case, chondrules are formed before parent body accretion, either by melting dust aggregates by nebular shocks \citep{2002M&PS...37..183D,2010ApJ...722.1474M,2016M&PS...51..870M}, for example related to global disk instabilities \citep{2005ApJ...621L.137B,lichtenberg15}, or condensation of melts and crystals \citep{2004M&PS...39.1897B,2008GeCoA..72.1442N}. In contrast, if chondrules formed via processes involving already formed planetesimals, the interpretation of their role would shift to a `by-product' of planet(esimal) formation (see discussion in Section \ref{sec:discussion}). 

Recently proposed chondrule formation scenarios considered melt spray from subsonic collisions (`splashes') between similar-sized planetesimals, which were fully melted by decay heat from $^{26}$Al \citep{2011E&PSL.308..369A,2012M&PS...47.2170S} or impact `jetting' via collisions of planetesimals with undifferentiated protoplanets \citep{2015Natur.517..339J,2016ApJ8168H,2017ApJ...834..125W}. Collisional mechanisms were suggested previously and offer attractive solutions to many chondrule features \citep{2005Natur.436..989K,2012M&PS...47.2170S,2014Icar..242....1S,2014ApJ...794...91D,2016ApJ...832...91D,2016SciA....2E1001M}. From a dynamical point-of-view, collisional interactions of planetesimals and embryos during accretion are inevitable and expected to create a vast amount of continuously reprocessed debris \citep{2006Natur.439..821B,2015ApJ...813...72C,2015GMS...212...49J,asphaug2017,bottke2017} that inherits the geochemical features from previous planetesimal generations.

Collisional models of chondrule formation considering fully-molten planetesimals, and thus highly energetic internal magma oceans with temperatures above the liquidus \citep{2011E&PSL.308..369A,2012M&PS...47.2170S}, have the advantage that bodies interacting at low speeds ($\sim$ around the two-body escape velocity) can cause a melt spray ejection into the ambient disk medium that provides the inferred cooling regime for chondrules and the required solid densities to preserve primitive abundances of moderately volatile elements \citep{2012M&PS...47.2170S,2014ApJ...794...91D,2016ApJ...832...91D}.

For consistency with the observed metal abundances in and around chondrules \citep{2010GeCoA..74.2212W,2014ChEG...74..507P,2016JGRE..121.1885C}, droplet entrainment in a vigorously convecting magma ocean has been invoked to prevent efficient and complete metal-silicate segregation \citep{2011E&PSL.308..369A,2012M&PS...47.2170S,asphaug2017}. However, metal sequestration into the planetesimal core may have been rapid in magma ocean planetesimals as, for instance, supported by the old ages of iron meteorites \citep{2014Sci...344.1150K}. In this case, re-establishing post-collisional bulk Fe/Mg ratios and forming chondrites with metal beads would require a complicated and highly unlikely scenario of (i) partial oxidization of the metal cores of fully differentiated planetesimals and (ii) violent remixing of the remaining metal core material with mantle silicates during or after the collision \citep{2015E&PSL.411...11P}. Additionally, chemical \citep{2005ASPC..341..251J,2007GeCoA..71.4092H,2014ChEG...74..507P} and isotopic \citep{2016LPICo1921.6503B,olsen2016magnesium} heterogeneities between single chondrules of the same meteorite cannot be retained if vigorous convection at low silicate viscosities homogenized the bulk volume of primitive planetesimals down to chondrule-sized microscales. 

However, it is well known that the interior evolution of planetesimals alone could create a diverse range of thermal histories and interior structures \citep[e.g.,][]{2006M&PS...41...95H,lichtenberg16a}, where magma ocean planetesimals are only one end-member type. In addition, the structure and chemistry of planetary materials was potentially further altered due to repeated collision--reaccretion cycles, which may generate varying thermal and chemical histories on a cm--m scale of planetary materials. 
Here, we probe the thermal and chemical evolution of such debris in a dynamical setting for the early solar system, where small ($<$ 100 km) planetesimals were continuously formed over a given timeframe during the lifetime of the circumstellar disk, evolved internally due to radiogenic heating, and were subsequently destroyed by collisions. To evaluate the thermal and chemical state of the debris over time, we quantify the processes governing metal-silicate segregation and chemical diversity within molten planetesimals and model their thermal histories dependent on their sizes and initial \al abundances. To classify the parameter space that is (in-)consistent with chondrule formation from impact splashes among similar-sized planetesimals, we calculate the combined influence of interior evolution and collisional parameters in a simple Monte Carlo model. We describe our methodology in Section \ref{sec:materials} and show the results from our scalings and computations in Section \ref{sec:results}. We discuss our findings and the limits of our approach in Section \ref{sec:discussion}, and draw conclusions in Section \ref{sec:conclusions}.

%%%%%%%%%%%%%%%%%%%%%%%%%%%%%%%%%%%%%%%%%%%%%%%%%%%%%%%%%%%%%%%%%%%%%%%%%%%%%%%%%%%%%%%%%%%%%%
%%%%%%%%%%%%%%%%%%%%%%%%%%%%%%%%%%%%%%%%%%%%%%%%%%%%%%%%%%%%%%%%%%%%%%%%%%%%%%%%%%%%%%%%%%%%%%
%%%%%%%%%%%%%%%%%%%%%%%%%%%%%%%%%%%%%%%%%%%%%%%%%%%%%%%%%%%%%%%%%%%%%%%%%%%%%%%%%%%%%%%%%%%%%%
%%%%%%%%%%%%%%%%%%%%%%%%%%%%%%%%%%%%%%%%%%%%%%%%%%%%%%%%%%%%%%%%%%%%%%%%%%%%%%%%%%%%%%%%%%%%%%
%%%%%%%%%%%%%%%%%%%%%%%%%%%%%%%%%%%%%%%%%%%%%%%%%%%%%%%%%%%%%%%%%%%%%%%%%%%%%%%%%%%%%%%%%%%%%%
%%%%%%%%%%%%%%%%%%%%%%%%%%%%%%%%%%%%%%% METHODS %%%%%%%%%%%%%%%%%%%%%%%%%%%%%%%%%%%%%%%%%%%%%%
%%%%%%%%%%%%%%%%%%%%%%%%%%%%%%%%%%%%%%%%%%%%%%%%%%%%%%%%%%%%%%%%%%%%%%%%%%%%%%%%%%%%%%%%%%%%%%
%%%%%%%%%%%%%%%%%%%%%%%%%%%%%%%%%%%%%%%%%%%%%%%%%%%%%%%%%%%%%%%%%%%%%%%%%%%%%%%%%%%%%%%%%%%%%%
%%%%%%%%%%%%%%%%%%%%%%%%%%%%%%%%%%%%%%%%%%%%%%%%%%%%%%%%%%%%%%%%%%%%%%%%%%%%%%%%%%%%%%%%%%%%%%
%%%%%%%%%%%%%%%%%%%%%%%%%%%%%%%%%%%%%%%%%%%%%%%%%%%%%%%%%%%%%%%%%%%%%%%%%%%%%%%%%%%%%%%%%%%%%%
%%%%%%%%%%%%%%%%%%%%%%%%%%%%%%%%%%%%%%%%%%%%%%%%%%%%%%%%%%%%%%%%%%%%%%%%%%%%%%%%%%%%%%%%%%%%%%
\section{Methods}
\label{sec:materials}

\subsection{Scaling analysis}

This first part of our analysis aims to quantify the thermochemical processes governing the interior of planetesimals with high melt fractions above the rheological transition. The rheological transition of silicates describes the critical melt fraction \phicrit $\sim$ 0.4--0.6 \citep{2009GGG....10.3010C} at which the silicate viscosity drops by orders of magnitude (from rock- to water-like behavior). At this range, the dynamic state of the system changes from solid-state creep processes to liquid-like convectional motions in an interior magma ocean. Here, we describe the processes in an idealized system that represents the end-member scenario of a planetesimal that has fully melted as a result of \al decay.

\subsubsection{Metal-silicate segregation from Fe,Ni droplet rainfall}
\label{sec:scaling1}

\begin{table*}[bth]
\centering
\caption{List of physical parameters used.}
\label{tab:constants}
\begin{tabular}{lllll}
\hline
\textsc{Parameter} & \textsc{Symbol} &  \textsc{Value} & \textsc{Unit} & \textsc{References} \\
\hline
Density of uncompressed solid silicates  & $\rho_{\mathrm{Si-sol}}$ & 3500 & kg m$^{-3}$ & (1,2) \\
Density of uncompressed molten silicates & $\rho_{\mathrm{Si-liq}}$ & 2900 & kg m$^{-3}$ & (1)  \\
Density of uncompressed iron & $\rho_{\mathrm{Fe}}$ & 7540 & kg m$^{-3}$ &  (3) \\
Ambient temperature & $T_{\mathrm{0}}$ & 290 & K & (3,4) \\
Activation energy & $E_{\mathrm{a}}$ & 470 & kJ mol$^{-1}$  & (5)  \\
Dislocation creep onset stress & $\sigma_{\mathrm{0}}$ & $3 \cdot 10^7$ & Pa & (6) \\
Power law exponent & $n$ & 4 & non-dim. & (5)  \\
Latent heat of silicate melting & $L_{\mathrm{Si}}$ & 400 & kJ kg$^{-1}$ & (3,6) \\
Melt fraction at rheological transition  & $\varphi_{\mathrm{crit}}$ & 0.4 & non-dim. & (8,9) \\
Silicate heat capacity & $c_{\mathrm{P}}$ & 1000 & J kg$^{-1}$ K$^{-1}$ & (6)   \\
Thermal diffusivity & $\kappa$ & $1 \cdot 10^{-6}$ & m$^2$ s$^{-1}$ & (6)  \\
Thermal expansivity of solid silicates & $\alpha_{\mathrm{Si-sol}}$ & $3 \cdot 10^{-5}$ & K$^{-1}$ & (2)  \\
Thermal expansivity of molten silicates & $\alpha_{\mathrm{Si-liq}}$ & $6 \cdot 10^{-5}$ & K$^{-1}$ & (2)   \\
Thermal expansivity of iron & $\alpha_{\mathrm{Fe}}$ & $1 \cdot 10^{-5}$ & K$^{-1}$ & (9)  \\
Thermal conductivity of solid silicates & $k_{\mathrm{}}$ & 3 & W m$^{-1}$ K$^{-1}$ & (10)   \\
Thermal conductivity of molten silicates & $k_{\mathrm{eff}}$ & $\le 10^6$ & W m$^{-1}$ K$^{-1}$ & (11)   \\
Min. unsintered silicate thermal conductivity   & $k_{\mathrm{low}}$ & $10^{-3}$ & W m$^{-1}$ K$^{-1}$ & (12,13)  \\
Temperature at onset of hot sintering & $T_{\mathrm{sint}}$ & 700 & K & (12)   \\
Peridotite solidus temperature & $T_{\mathrm{sol}}$ & 1416 & K & (14)  \\
Peridotite liquidus temperature & $T_{\mathrm{liq}}$ & 1973 & K & (15)  \\
Lower cut-off viscosity & $\eta_{\mathrm{num}}$ & $10^{17}$ & Pa s & (16)  \\
Silicate melt viscosity & $\eta_{\mathrm{melt}}$ & $10^{-2}$ & Pa s & (17)  \\
Droplet surface energy & $\sigma$ & 1 & N m$^{-1}$ & (18)  \\
Chemical diffusivity of silicates & $\kappa_{\mathrm{C}}$ & $10^{-8}$ & m$^{2}$ s$^{-1}$ & (18)  \\
\hline
\end{tabular}
\\
\footnotesize{
References: (1) \citet{1981JGR....86.6261S}, (2) \citet{1998PEPI..107...53S}, (3) \citet{1998Icar..134..187G}, (4) \citet{1976ARAA..14...81B}, (5) \citet{ranalli1995rheology}, (6) \citet{turcotte2014geodynamics}, (7) \citet{2009GGG....10.3010C}, (8) \citet{solomatov2015magma}, (9) \citet{1990JGR....9521731B}, (10) \citet{2012Sci...338..939T}, (11) \citet{2014MPS...49.1083G}, (12) \citet{1984EPSL..68...34Y}, (13) \citet{2012AA...545A.135H}, (14) \citet{2000GGG.....1.1051H}, (15) \citet{2002EPSL.197..117T}, (16) \citet{lichtenberg16a}, (17) \citet{2005EPSL.240..589L}, (18) \citet{2003EPSL.205..239R}} 
\end{table*}

For the case of a fully-molten planetesimal, we parameterize the rain-out of Fe,Ni metal droplets following the description by \citet{solomatov2015magma}. The dynamic processes in the magma ocean are determined by its viscosity, which drops by orders of magnitude at the rheological transition \phicrit $\sim$ 0.4--0.6 \citep{2009GGG....10.3010C}, from $\eta \sim 10^{17}$ Pa s to $10^{-2}$ Pa s \citep{2003EPSL.205..239R, 2005EPSL.240..589L}, as listed in Table \ref{tab:constants}. In melt regimes valid for planetesimals, the convective heat flux $q$ of the magma ocean can be calculated via
\begin{eqnarray}
q = 0.089 k \frac{(T_{\mathrm{m}}-T_{\mathrm{0}}) Ra^{1/3}}{D},
\end{eqnarray}
with Rayleigh number
\begin{eqnarray}
Ra = \alpha_{\mathrm{Si-liq}} g \rho_{\mathrm{ref}} \frac{(T_{\mathrm{m}}-T_{\mathrm{0}}) D^3}{\kappa \eta}, 
\end{eqnarray} 
potential temperature $T_{\mathrm{m}}$, ambient (and surface) temperature $T_{\mathrm{0}} = 290$ K, thermal diffusivity $\kappa = k/(\rho c_{\mathrm{p}})$, thermal conductivity of solid silicates $k$, silicate heat capacity $c_{\mathrm{p}}$, thermal expansivity of molten silicates $\alpha_{\mathrm{Si-liq}}$, depth of the magma ocean $D$, silicate densities
\begin{eqnarray}
\rho_{\mathrm{s}} & = &\rho_{\mathrm{sol}} - ( \rho_{\mathrm{sol}}-\rho_{\mathrm{liq}} ) \cdot \varphi,\\
\rho_{\mathrm{sol}} & = & \rho_{\mathrm{Si-sol}} \cdot (1 - \alpha_{\mathrm{Si-sol}} \cdot [T-T_{\mathrm{0}}]),\\
\rho_{\mathrm{liq}} & = & \rho_{\mathrm{Si-liq}} \cdot (1 - \alpha_{\mathrm{Si-liq}} \cdot [T-T_{\mathrm{0}}]),
\end{eqnarray}
temperature $T$ and thermal expansivity of solid silicates $\alpha_{\mathrm{Si-sol}}$. 
See Table \ref{tab:constants} for the numerical values used. The convective velocities are then
\begin{eqnarray}
v_{\mathrm{s}} \approx 0.6 \left( \frac{\alpha_{\mathrm{Si-liq}} g l q}{\rho_{\mathrm{s}} c_{\mathrm{p}}} \right) ^{1/3}, \label{eq:vs}
\end{eqnarray}
with gravity $g$, and mixing length $l \sim D \sim$ \RP. Based on laboratory experiments, it has been shown that droplets can be suspended (or re-entrained) by the convective flow if their diameter is
\begin{eqnarray}
d \leq \frac{\rho_{\mathrm{s}} (v_{\mathrm{s}}/x_{*})^2}{0.1 ( \rho_{\mathrm{m}} - \rho_{\mathrm{s}}) g}, \label{eq:suspension}
\end{eqnarray}
with metal density
\begin{eqnarray}
\rho_{\mathrm{m}} = \rho_{\mathrm{Fe}} \cdot (1 - \alpha_{\mathrm{Fe}} \cdot [T-T_{\mathrm{0}}]),
\end{eqnarray}
thermal expansivity of iron $\alpha_{\mathrm{Fe}}$ and constant factor $x_{*} = 60$ \citep{solomatov2015magma}. Metal droplets suspended in the magma tend to be drawn together into spherical droplets, minimizing their surface area. Their stability is determined by the ratio between the stagnation pressure and the internal pressure caused by surface tension, given by the Weber number $We$, which can be used to estimate the expected sizes of droplet diameters
\begin{eqnarray}
d = \frac{\sigma \cdot We}{(\rho_{\mathrm{m}} - \rho_{\mathrm{s}}) v_{\mathrm{s}}^2}, \label{eq:stability}
\end{eqnarray}
with surface energy $\sigma$, where $We \leq 10$ is the stability threshold \citep{2003EPSL.205..239R}. For a given depth of the magma ocean $D \sim$ \RP, melt fraction $\varphi$, and the numerical values listed in Table \ref{tab:constants}, we can determine the ratio of the expected droplet sizes and the upper limits for suspension (Section \ref{sec:results1}, Figure \ref{fig:1}). The expected droplet size must be smaller than the upper limit for suspension for the droplets to be entrained in the flow and resist rain-out onto the planetesimal core.

\subsubsection{Chemical equilibration via turbulent diffusion}
\label{sec:coll_vs_equ}

Metal-silicate separation via the rainfall mechanism is not the only process that shapes the interior dynamics of a fully-molten planetesimal. Chemical and nucleosynthetic heterogeneities, inherited from the solar nebula prior to planetesimal formation, can be erased by large-scale convective mixing once the silicate rheology transitions to fluid-like behaviour. In addition to planetesimal-scale mixing, chemical diffusion \citep{2003EPSL.205..239R} from dissipation of turbulent energy down to the so-called Kolmogorov microscales affects chondrule- and grain-sized regions -- the precursor material for chondrules in the splashing model. The time scale at which neighbouring cells achieve such miscroscale chemical equilibration can be estimated via the local diffusion time scale
\begin{eqnarray}
t\rmu{eq} = l\rmu{K}^2/\kappa\rmu{C}, \label{eq:equilibration_time}
\end{eqnarray}
with the Kolmogorov length scale $l\rmu{K}$ and the chemical diffusivity $\kappa\rmu{C}$ (Table \ref{tab:constants}). The Kolmogorov length scale is given as 
\begin{eqnarray}
l\rmu{K} = \left( \frac{\eta\rmu{melt}^3 D}{\rho\rmu{Si-liq}^3 v\rmu{s}^3} \right)^{1/4},
\end{eqnarray}
with the viscosity of the magma ocean $\eta\rmu{melt}$, the density of molten silicates $\rho\rmu{Si-liq}$ and $v\rmu{s}$ the convection velocity in the magma ocean calculated from Equation \ref{eq:vs}. Like in the section before, we choose to approximate the magma ocean length scale $D$ with the planetesimal radius \RP, because in turbulent systems the dissipation rate at the smallest scales is primarily determined by the length scale of total kinetic energy in the turbulent motions; that is, the planetesimal radius in the case of a fully-molten magma ocean planetesimal. Using these scalings, we compute the time scales for diffusion, dependent on the radius of the planetesimal and the silicate viscosity.

In addition, we plot the time scales for collisions between similar-sized objects \citep{2010ChEG70199A} in a planetesimal collision setting using
\begin{eqnarray}
t\rmu{coll} = \frac{2 R\rmu{P}}{\Delta v} \label{eq:collision_time}
\end{eqnarray}
with the impact velocity $\Delta v$. Using the parameters given in Table \ref{tab:constants}, we derive scalings for various melt fractions, planetesimal radii and impact velocities, which are displayed in Figure \ref{fig:S1} (Section \ref{sec:results1}). 

\subsection{Thermomechanical evolution of planetesimals}
\label{sec:thermomodels}

\begin{figure}[tb]
    \centering
    \includegraphics[width=0.32\textwidth]{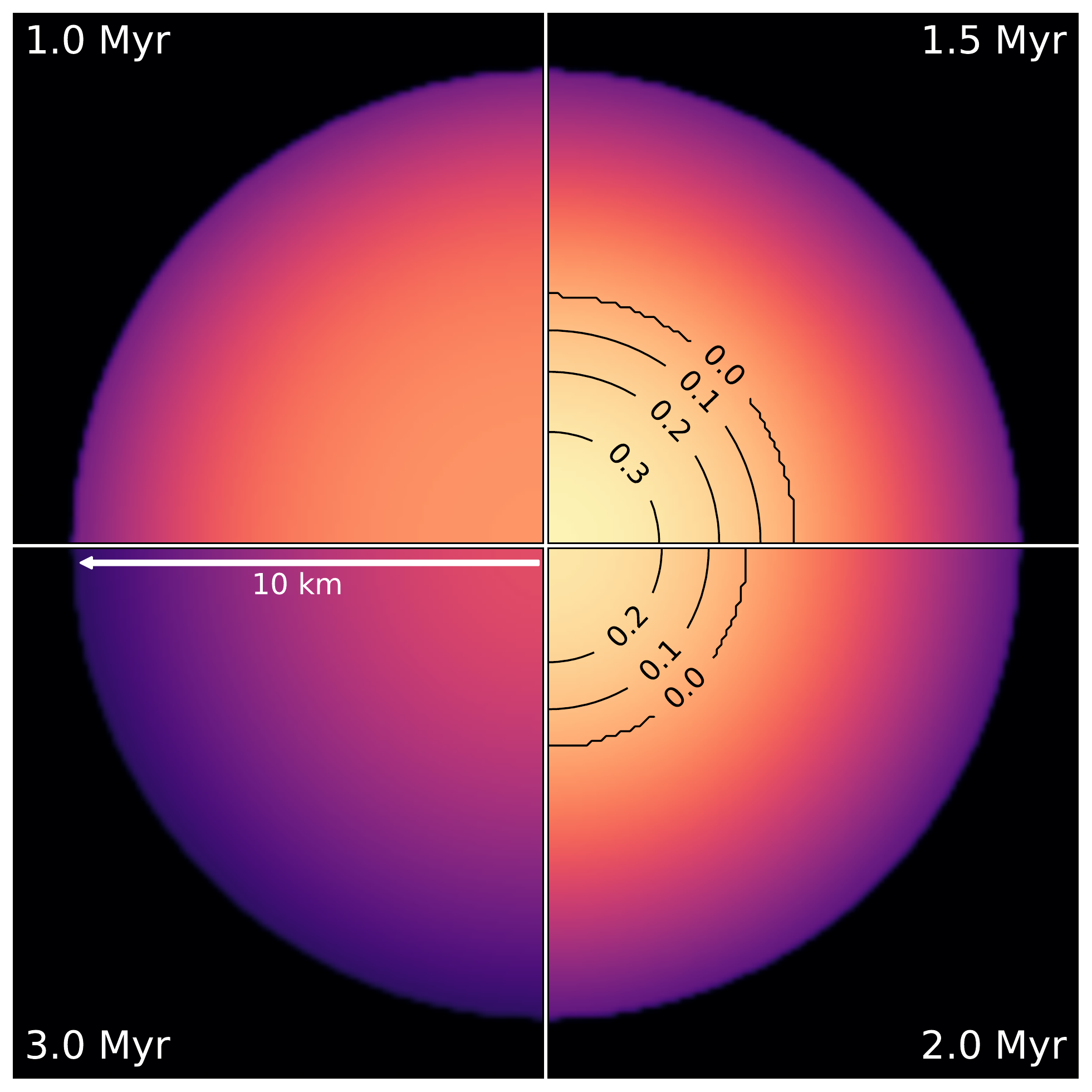}
    \caption{Thermal evolution of a planetesimal that is 10 km in radius and formed at 0.5 Myr after CAIs in our 2D cylinder geometry models. Silicate melt fractions are indicated with black isolines, and the temperature color scale ranges linearly from $\sim$290 K (black) to $\sim$1650 K (bright yellow).}
    \label{fig:9}
\end{figure}

\begin{figure}[tbh]
    \centering
    \includegraphics[width=0.49\textwidth]{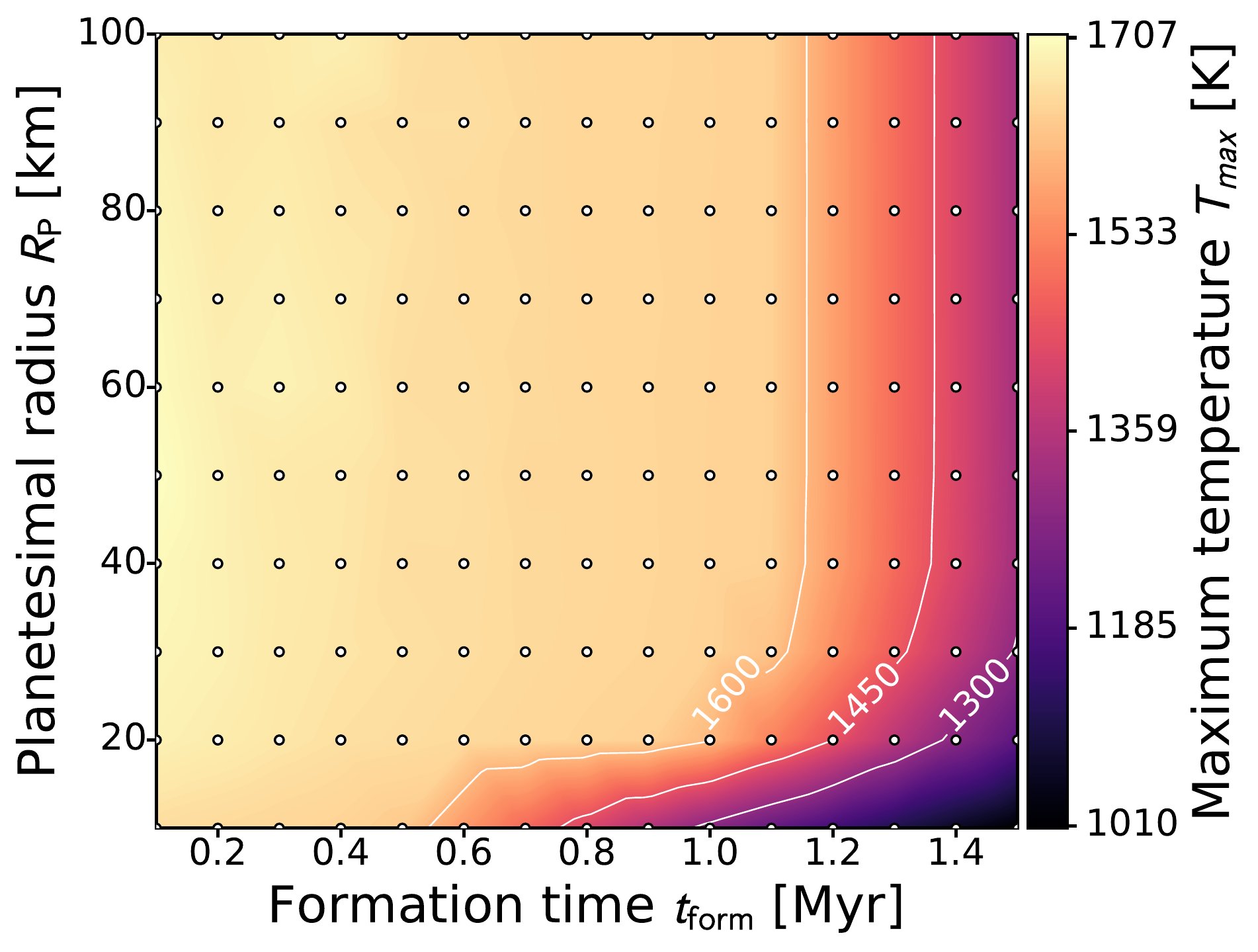}
    \caption{Maximum temperatures $T_{\mathrm{max}}$ reached within planetesimals of radius \RP and formation time \tform. Small planetesimals below 20--30 km in radius reach significantly lower temperatures than larger bodies for a given formation time.}
    \label{fig:10}
\end{figure}

To bring the former calculations into context, we now consider the time-dependent interior evolution of planetesimals participating in potential chondrule-forming collisions in the early solar system. To do so, we model their thermomechanical histories using two-dimensional fluid dynamics simulations employing a conservative finite-differences fully-staggered grid formulation \citep{2003PEPI..140..293G,2007PEPI..163...83G}. The numerical model is described in detail in \citet{2014MPS...49.1083G} and \citet[][and references therein]{lichtenberg16a}, which is why we only briefly summarize its main characteristics here. The code solves the continuity, Stokes and location-dependent Poisson equation for self-gravity of material together with the energy equation, which includes source terms for radiogenic, shear and latent heat production. Physical properties are advanced using Lagrangian markers to minimize numerical diffusion and capture sharp viscosity and temperature gradients. To account for the solar system-specific $^{26}$Al heating term \citep{2016MNRAS.462.3979L}, the so-called `canonical' abundance of $^{26}$Al/$^{27}$Al = $5.25 \cdot 10^{-5}$ at CAI formation \citep{2013M&PS...48.1383K} is adopted.

The silicate melt is parameterized according to a peridotitic composition, taking into account both consumption and release of latent heat. For melt fractions $\varphi \geq$ 0.4 the convective heat flux is approximated using the soft turbulence formulation \citep{1962PhFl....5.1374K,1994AnRFM..26..137S}. All our models incorporate an initial macroporosity (inverse filling-factor) of $\phi\rmu{init}$ = 0.3, where sintering and compaction effects are parameterized using constraints from laboratory experiments \citep{2012AA...545A.135H,2015AA...576A..60G}.

The numerical models were run using a two-dimensional infinite cylinder geometry on a Cartesian grid, starting from ambient temperature of $T_{\mathrm{0}}$ = 290 K, as for such small bodies accretionary heat is insignificant \citep{schubert1986thermal,2011EPSL.305....1E}, and are surrounded by a low-density and low-viscosity layer of 'sticky air' that serves as heat sink \citep{2008PEPI..171..198S,2012GeoJI.189...38C}. The parameter space investigated spans the regime of \RP = 10--100 km in steps of 10 km, and \tform = 0.1--1.5 Myr after CAIs in steps of 0.1 Myr, the potential formation time interval of chondrule precursor material \citep{2015PNAS..112.1298L}. Illustrations of the two-dimensional temperature and melt fraction evolution for a single simulation and the entire simulation grid are shown in Figure \ref{fig:9} and Figure \ref{fig:10}, respectively. Further visualizations and analyses of the major qualitative regimes of the time-dependent thermal and density evolution are shown in \citet{lichtenberg16a}.

Importantly, our model utilizes a scaling for the cooling of a low-viscosity magma ocean, in which the effective thermal conductivity across finite-difference nodes is given as
\begin{eqnarray}
k\rmu{eff} = \left( q/0.089 \right)^{3/2} \cdot \frac{\alpha_{\mathrm{Si-liq}} g c\rmu{p}}{\Delta T^2 \rho\rmu{s} \eta\rmu{num}}
\end{eqnarray}
with the convective heat flux $q$, the temperature difference across the nodes $\Delta T$, the silicate density $\rho\rmu{s}$, the thermal expansivity $\alpha_{\mathrm{Si-liq}}$, gravity $g$, and the lower cut-off viscosity $\eta\rmu{num} = 10^{17}$ Pa s. This effective heat flux numerically approximates the increased heat flux during magma ocean stages and results in a more effective cooling of regions which are subject to the highest temperatures \citep{2006M&PS...41...95H,lichtenberg16a}. We use these interior evolution models to determine the time-dependent thermal structure of the planetesimals together with the scalings from Section \ref{sec:scaling1} and \ref{sec:coll_vs_equ} to evaluate which parts of their interior  can be eligible as chondrule precursor bodies at a given time after the formation of CAIs.

\subsection{Evolution-collision model}
\label{sec:montecarlo}

The evolution of the precursor bodies is important to understand the energetic state and evolution of the silicates before the chondrule forming impact event. However, if the planetesimal body is not \emph{fully} molten ($\varphi <$ \phicrit), the impact energy is necessary to elevate the silicate temperatures to above the chondrule formation temperature of \Tchondrule $\geq$ 1900 K and eject the molten material from the two bodies. When material is freed from the colliding bodies during the impact, it is first compressed by the impact shock wave and then decompressed after ejection. By this process, fractions of the lithostatic/hydrostatic pressure within the planetesimal are converted into surface energy of magma (chondrule) droplets \citep{2011E&PSL.308..369A,asphaug2017}. In order to produce a melt spray that is consistent with the thermal histories of chondrules, at least parts of the material must have been heated to \Tchondrule $\geq$ 1900 K \citep{2008Sci...320.1617A,2016JGRE..121.1885C} and subsequently cool down in an emerging droplet cloud of high density \citep{2014ApJ...794...91D,2016ApJ...832...91D}.

In order to demonstrate the influence of the pre-collision state on chondrule thermal histories and the post-collision energy distribution, we have developed a Monte Carlo approach. The {\em a priori} assumption for this model is that planetesimals collide continuously during 0--5 Myr after CAI formation. Furthermore, when bodies collide, they generate debris and new planetesimals may be created from this material. For the moment, bodies from primordial and reaccreted material are treated the same, i.e., the material does not have a chemical `memory' of prior generations. We discuss these and other assumptions in Section \ref{sec:discussion}.

We start by randomly generating planetesimals in agreement with a radius power law
\begin{eqnarray}
\mathrm{d}N/\mathrm{d}R_{\mathrm{P}} \propto R_{\mathrm{P}}^{-q},
\end{eqnarray}
with the number of bodies $N$ and power law index $q = 2.8$, consistent with shearing-box simulations of the streaming instability mechanism \citep{2015SciA....115109J, 2016ApJ...822...55S,2017arXiv170503889S}. Using this power law, we generate integer radii \RP $\geq R_{\mathrm{P,min}}$ according to
\begin{eqnarray}
R_{\mathrm{P}} = ||R_{\mathrm{P,min}} (1-x_{\mathrm{rand}})^{-1/(q-1)}||,
\end{eqnarray}
with the minimum planetesimal radius in our parameter space $R_{\mathrm{P,min}}$ = 10 km and pseudo-random number $x_{\mathrm{rand}}$ = 0--1. Depending on the regime chosen ($R_{\mathrm{P,max}}$ = 20, 30, 50, 100 km) we accept or reject radii exceeding the upper limit value, resulting in an approximate power law distribution.

Following the approach of \citet{1993Icar..106..190W} \citep[as described in][Supplementary Material therein]{2009Icar..204..558M}, we build a normalized collision probability distribution of pair-encounters for the generated planetesimals during a time step $\delta t$ using 
\begin{eqnarray}
\hat{N}_{\mathrm{c,ij}} = N_{\mathrm{i}} N_{\mathrm{j}} F_{\mathrm{g,ij}} (R_{\mathrm{i}} + R_{\mathrm{j}})^2,
\end{eqnarray}
with bodies of different sizes $i$ and $j$ with their respective numbers $N\rmu{i}$ and $N\rmu{j}$ and radii $R\rmu{i}$ and $R\rmu{j}$ and gravitational focusing factor $F_{\mathrm{g,ij}} \approx 1$ for the velocity dispersions chosen here. Next, we sample the collision probability distribution using a linear alias method to return $N\rmu{P,tot}/2$ collision pairs $ij$, where $N\rmu{P,tot}$ is the total number of bodies in the generated planetesimal family. 

Each planetesimal in each collision pair is randomly assigned a formation time \tform = [\tmin, \tmax], with \tmin = 0.1 or 0.5 Myr, and \tmax = 1.5 Myr. Additionally, we investigated a parameter space where \tmin and \tmax varied with collision time \tcollision, such that \tmin = \tcollision - $\Delta t$, where $\Delta t$ = 0.5 or 0.7 Myr (but \tmin = 0.1 Myr at minimum and \tmin = 1.0/0.8 Myr at maximum) and \tmax = 1.5 Myr. Naturally, the maximum formation time was limited to \tcollision in case \tcollision $< t\rmu{form,max}$ = 1.5 Myr. The collision pair is assigned a randomized impact angle $\theta$ = [35, 55]$^{\circ}$ and a collision velocity $\Delta v$ = 0.5, 1.0, 1.5 or 2.0 km/s, according to the specific setting. The different parameter choices are summarized in Table \ref{tab:montecarlo}.

\begin{table}[tbh]
\centering
\caption{Parameters for the Monte Carlo collision model with the maximum radius of a planetesimal family $R_{\mathrm{P,max}}$, the earliest formation $t\rmu{form,min}$, the latest formation $t\rmu{form,max}$, the maximum planetesimal dwell time $\Delta t$ and the velocity dispersion $\Delta v$. See text for details on the assumptions.}
\label{tab:montecarlo}
\begin{tabular}{lll}
\hline
Parameter & Unit &  Values \\
\hline
 $R_{\mathrm{P,max}}$ & km & 20, 30, 50, 100 \\
 $t\rmu{form,min}$ & Myr & 0.1, 0.5 \\
 $t\rmu{form,max}$ & Myr & 1.5 \\
 $\Delta t$ & Myr & 0.5, 0.7 \\
 $\Delta v$ & km/s & 0.5, 1.0, 1.5, 2.0\\
\hline
\end{tabular}
\end{table}

We evaluate each collision, depending on the sizes (and therefore masses, assuming constant initial densities of $\rho$ = 3500 kg/m$^3$) $R\rmu{i}$ and $R\rmu{j}$, angle $\theta$ and impact speed $\Delta v$, employing the \textsc{edacm} model of \citet{2012ApJ74579L}. If the outcome is super-catastrophic, defined as the mass of the largest intact remnant block after the collision being less than 0.1 of the combined mass, we calculate the thermal effect on the remnant material as described further below. We note that the model of \citet{2012ApJ74579L} has recently been challenged regarding the catastrophic disruption threshold. \citet{2016Icar27585M} argue for a lower threshold value than \citet{2012ApJ74579L}, thus our estimate for super-catastrophic break-up using the \textsc{edacm} scaling can be seen as a conservative approach so that we do not overestimate the number of super-catastrophic collisions, and thus potential chondrule material.

For categorizing the collisional debris, we evaluate the thermal state and material properties of each planetesimal in a collision pair before the impact using a two-dimensional bilinear interpolation from the initial \RP-\tform parameter grid (see Figure \ref{fig:10} and Section \ref{sec:results1}) of the time-dependent numerical models described in Section \ref{sec:thermomodels}. The injected energy from body $i$ to $j$, $\Delta E\rmu{ij} = E\rmu{kin,i} - E\rmu{pot,j}$, with the kinetic energy of impactor $i$, $E\rmu{kin,i}$, and potential energy of target $j$, $E\rmu{pot,j}$, is homogenized over the target volume via $\Delta E\rmu{j,k} = (E\rmu{kin,k}/E\rmu{pot,j})\Delta E\rmu{ij}$, where the target volume is sub-divided into $n$ shells with energy $E\rmu{pot,k}$. If the injected energy $\Delta E\rmu{j,k}$ into a sub-volume of the target is greater than the energy needed to heat it to above the chondrule formation temperature \Tchondrule, the material is categorized as post-collision liquid $\hat{V}_{\mathrm{pc,chondrule}}$ (\Tpost $>$ \Tchondrule, Figure \ref{fig:thermal1}), normalized by the total debris volume of the colliding family of planetesimals. The necessary energy is given by 
\begin{eqnarray}
\Delta E_{\mathrm{chondrule,k}} = & & [ ( T_{\mathrm{chondrule}} - T_{\mathrm{k}} ) \cdot c_{\mathrm{p}}
\\
& + & L_{\mathrm{Si}} \cdot ( \varphi_{\mathrm{chondrule}} - \varphi_{\mathrm{k}} ) ] \cdot m_{\mathrm{k}}, \nonumber
\end{eqnarray} 
with temperature $T\rmu{k}$, minimal chondrule formation peak melt fraction $\varphi\rmu{chondrule}$, latent heat of silicate melt $L\rmu{Si}$, melt fraction $\varphi\rmu{k}$ and mass $m\rmu{k}$ of the specific sub-volume. If it does not reach \Tchondrule, it is counted as $\hat{V}_{\mathrm{pc,residual}}$, further subdivided into partially melted (\Tchondrule $>$ \Tpost $>$ \Tsol, Figure \ref{fig:thermal1}) and unmelted material (\Tpost $<$ \Tsol, Figure \ref{fig:thermal1}). If the material exceeded the defined melt fraction threshold for metal loss \phicrit (0.4, 0.5 or 0.6) before the collision, the sub-volume is counted as $\hat{V}_{\mathrm{pc,loss}}$ ($\varphi\rmu{pre}$ $>$ \phicrit, Figure \ref{fig:thermal1}). We re-do these steps for each planetesimal of each collision pair for all timesteps starting from 0.1 Myr (or 0.5 Myr, \nameref{sec:suppl}) after CAI until 5 Myr after CAI, where for each timestep a new 'collision family' is generated. For instance, for model setting \tform = [0.1, 1.5] Myr at time $t$ = 0.8 Myr, the planetesimals in the colliding family were randomly drawn from a formation time interval \tform = [0.1, 0.8] Myr; at time $t$ = 2.3 Myr from a time interval \tform = [0.1, 1.5] Myr. For model setting \tform = [\tcollision - 0.5, 1.5] Myr at time $t$ = 0.8 Myr, the planetesimals in the colliding family were randomly drawn from a formation time interval \tform = [0.3, 0.8] Myr; at time $t$ = 2.3 Myr from a time interval \tform = [1.0, 1.5] Myr.

The approach outlined above has several simplifications. First, the intrinsic collision probability for bodies in the sampled planetesimal orbit is chosen to be 
\begin{eqnarray}
P_{\mathrm{ij}} = (\alpha_{\mathrm{v}} v_{\mathrm{ij}})/[4 H a (\delta a + 2 a e_{\mathrm{i}})] = \mathrm{const.},
\end{eqnarray} 
with the average collision velocity $\Delta v$, a constant depending on $\Delta v$ ranging from $0.57 \leq \alpha\rmu{v} \leq 0.855$ \citep{1993Icar..106..190W}, the symmetrical mutual scale height $H$, the semi-major axis and width $a$ and $\delta a$ of the annulus and the mean eccentricity of projectiles $e\rmu{i}$, is chosen to be constant. Therefore, we do not simulate a global source system of generated planetesimals that collide randomly. Rather, we \textit{ab initio} assume planetesimals that formed according to the power law slope described above and collide in pairs with a probability given by the mutual geometric factor. In other words, pairs of massive planetesimals are favored due to their larger geometrical cross-section, but eventually all planetesimals generated from the SFD do collide. Second, we consider only the simple cases of super-catastrophic interactions. In fact, catastrophic, hit-and-run, erosive and accretionary interactions could have an influence as well \citep{2010ChEG70199A}. However, for the low-mass regimes coupled with the chosen impact velocities shown here, super-catastrophic or catastrophic impacts are important and may create the majority of the debris. Third, the injected energy $\Delta E\rmu{ij}$ is assumed to fully go into disruption and heating energy of the target material.

We use this model to demonstrate the qualitative imprint of the pre-collision interior evolution state of the planetesimals on the collisional debris in Section \ref{sec:results2} and discuss its implications in Section \ref{sec:discussion}.

%%%%%%%%%%%%%%%%%%%%%%%%%%%%%%%%%%%%%%%%%%%%%%%%%%%%%%%%%%%%%%%%%%%%%%%%%%%%%%%%%%%%%%%%%%%%%%
%%%%%%%%%%%%%%%%%%%%%%%%%%%%%%%%%%%%%%%%%%%%%%%%%%%%%%%%%%%%%%%%%%%%%%%%%%%%%%%%%%%%%%%%%%%%%%
%%%%%%%%%%%%%%%%%%%%%%%%%%%%%%%%%%%%%%%%%%%%%%%%%%%%%%%%%%%%%%%%%%%%%%%%%%%%%%%%%%%%%%%%%%%%%%
%%%%%%%%%%%%%%%%%%%%%%%%%%%%%%%%%%%%%%%%%%%%%%%%%%%%%%%%%%%%%%%%%%%%%%%%%%%%%%%%%%%%%%%%%%%%%%
%%%%%%%%%%%%%%%%%%%%%%%%%%%%%%%%%%%%%%%%%%%%%%%%%%%%%%%%%%%%%%%%%%%%%%%%%%%%%%%%%%%%%%%%%%%%%%
%%%%%%%%%%%%%%%%%%%%%%%%%%%%%%%%%%%%%%% RESULTS %%%%%%%%%%%%%%%%%%%%%%%%%%%%%%%%%%%%%%%%%%%%%%
%%%%%%%%%%%%%%%%%%%%%%%%%%%%%%%%%%%%%%%%%%%%%%%%%%%%%%%%%%%%%%%%%%%%%%%%%%%%%%%%%%%%%%%%%%%%%%
%%%%%%%%%%%%%%%%%%%%%%%%%%%%%%%%%%%%%%%%%%%%%%%%%%%%%%%%%%%%%%%%%%%%%%%%%%%%%%%%%%%%%%%%%%%%%%
%%%%%%%%%%%%%%%%%%%%%%%%%%%%%%%%%%%%%%%%%%%%%%%%%%%%%%%%%%%%%%%%%%%%%%%%%%%%%%%%%%%%%%%%%%%%%%
%%%%%%%%%%%%%%%%%%%%%%%%%%%%%%%%%%%%%%%%%%%%%%%%%%%%%%%%%%%%%%%%%%%%%%%%%%%%%%%%%%%%%%%%%%%%%%
%%%%%%%%%%%%%%%%%%%%%%%%%%%%%%%%%%%%%%%%%%%%%%%%%%%%%%%%%%%%%%%%%%%%%%%%%%%%%%%%%%%%%%%%%%%%%%

\section{Results}
\label{sec:results}

In this section we present the results of our models of the thermomechanical history of colliding bodies before the impact event (Section \ref{sec:results1}) and the outcome of the coupled evolution-collision scenario (Section \ref{sec:results2}).

\subsection{Thermo-mechanical-chemical evolution before the collision event}
\label{sec:results1}

\begin{figure}[tbh]
    \centering
    \includegraphics[width=0.46\textwidth]{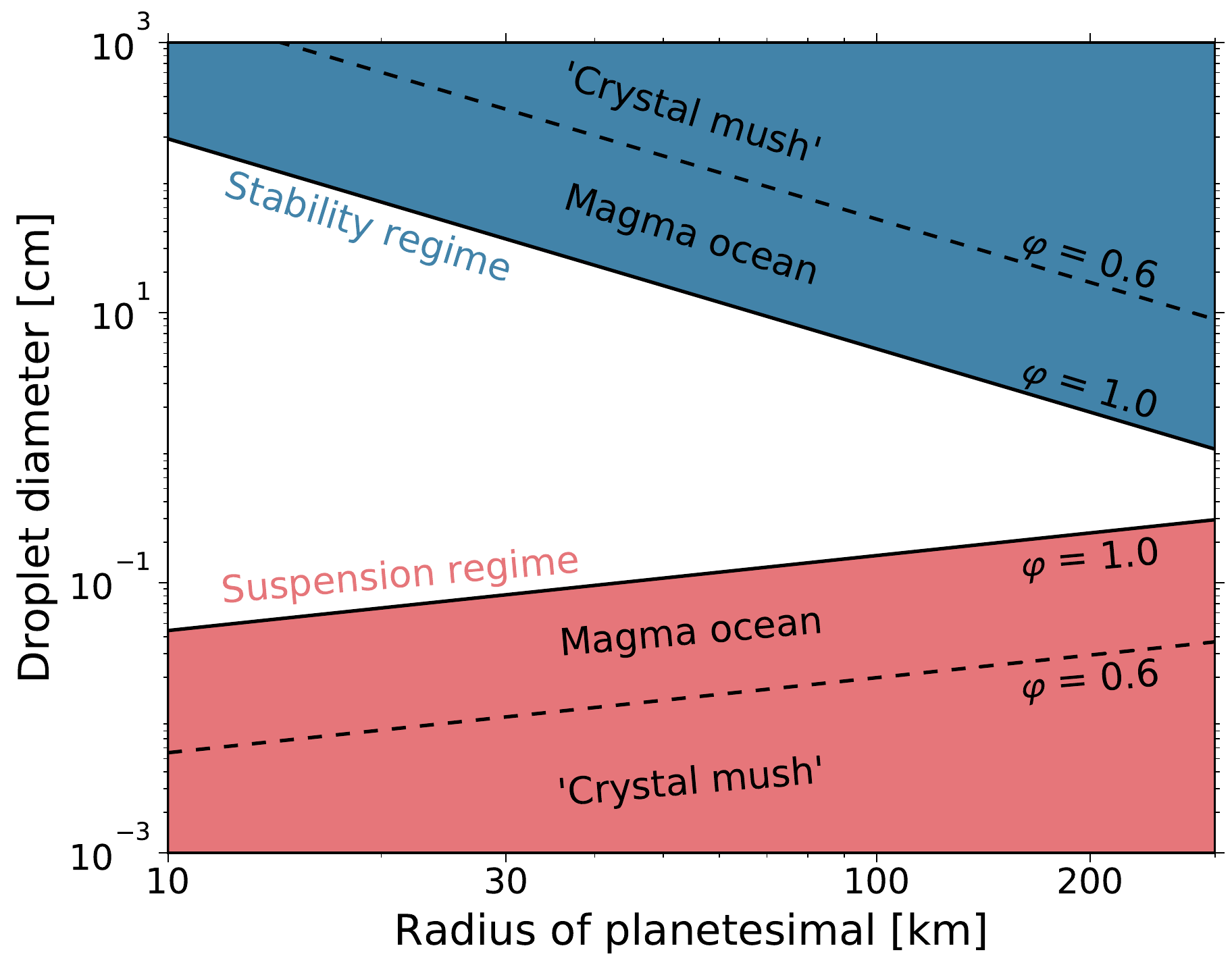}
    \caption{Droplet sizes of Fe,Ni metal versus planetesimal radius. The blue region ('stability'; Equation \ref{eq:stability}) shows the expected droplet sizes in fully-molten magma ocean planetesimals. The red region ('suspension'; Equation \ref{eq:suspension}), in contrast, shows the maximum droplet sizes that can be entrained by vigorous convection for various melt fractions $\varphi$. Since the suspension limit never exceeds the stability criterion, metal droplets in fully-molten planetesimals efficiently segregate into the core. See text for details on the scalings. The considered planetesimal radius range here, and in Figure \ref{fig:S1}, corresponds to the birth-size frequency distribution suggested by \citet{2015SciA....115109J}. }
    \label{fig:1}
\end{figure}

\begin{figure}[tbh]
  \centering
  \includegraphics[width=0.45\textwidth]{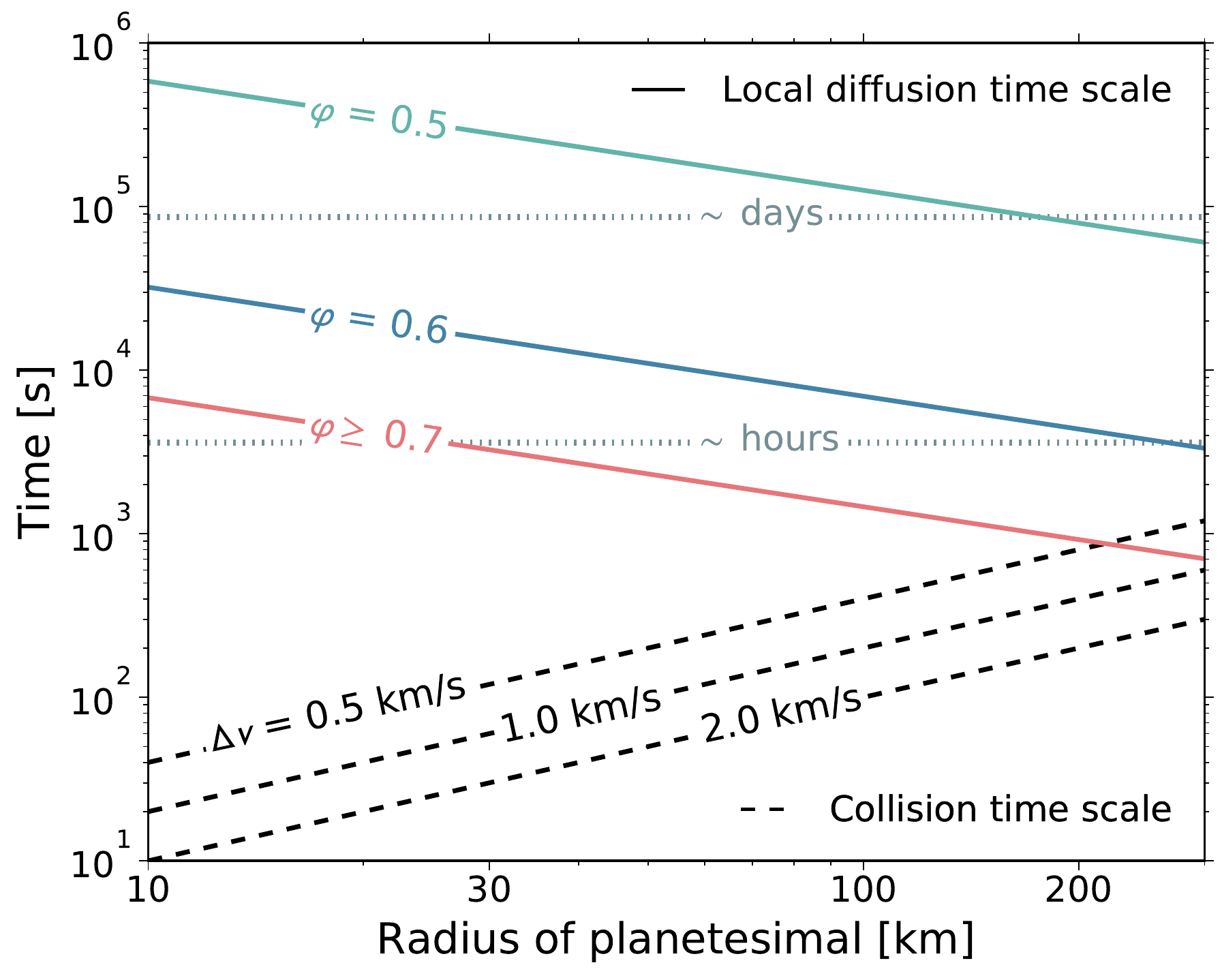}
  \caption{Radius of planetesimal versus time for either chemical equilibration (solid lines) or a planetesimal collision event (dashed lines). The solid lines ('local diffusion time scale', Equation \ref{eq:equilibration_time}) show the time scale for microscale chemical equilibration in a fully-molten magma ocean planetesimal for various silicate melt fractions $\varphi$. These are on the order of hours to days and thus demonstrate that fully-molten planetesimals rapidly chemically equilibrated. Therefore, any collisinal debris from them would feature chemical signatures {\em unlike} chondrite material. The dashed lines ('collision time scale', Equation \ref{eq:collision_time}), in contrast, quantify the time it takes for an average collision of two similar-sized planetesimals to take place at various encounter velocities $\Delta v$. Up to several hundreds of km in radius, the diffusion (solid) and the collision (dashed) time scales differ by orders of magnitude. That means, if the planetesimal material was chemically unequilibrated (= {\em not fully-molten}) before a hypothetical impact splash event, the expanding magma plume could retain a chemically and isotopically heterogeneous signature -- consistent with chondritic materials.}
  \label{fig:S1}
\end{figure}

Figures \ref{fig:1} and \ref{fig:S1} show the results of our scaling analysis of fully-molten planetesimals. Figure \ref{fig:1} demonstrates that the characteristic sizes of Fe,Ni metal droplets for the expected dynamics in a planetesimal magma ocean do not allow for droplet suspension. The droplets grow to sizes larger than can be suspended by convection and will thus rapidly rain out onto the planetesimal center. Therefore, fully-molten planetesimals rapidly evolved into a physically differentiated structure.

Figure \ref{fig:S1} shows, first, that the time scale for chemical equilibration ($\sim$ hours to days) suggests a very fast homogenization of the material during magma ocean stages. In particular, it is much shorter than the lifetime of the protoplanetary disk \citep[$\sim 3$--$5 \cdot 10^6$ yr,][]{2014prpl.conf..475A}, the thermo-mechanical evolution of planetesimal interiors \citep[$\sim$ 10$^5$--10$^6$ yr,][]{2006M&PS...41...95H} and the collisional evolution of an accreting planetesimal swarm \citep[$\sim$ 10$^4$--10$^5$ yr,][]{1993Icar..106..190W}. Second, the chemical equilibration time scales for the cases we consider in this manuscript lie orders of magnitudes above the characteristic collision time scales. This suggests that the primordial chemical and isotopic heterogeneities inherited from prior to accretion were homogenized rapidly after reaching the magma ocean stage. However, the equilibration time scale is not fast enough to homogenize the interior during the collision if it remained below the rheological transition before the event, since the diffusion time scale is longer than the collision time scale by orders of magnitude.

\begin{figure}[tbh]
    \centering
    \includegraphics[width=0.49\textwidth]{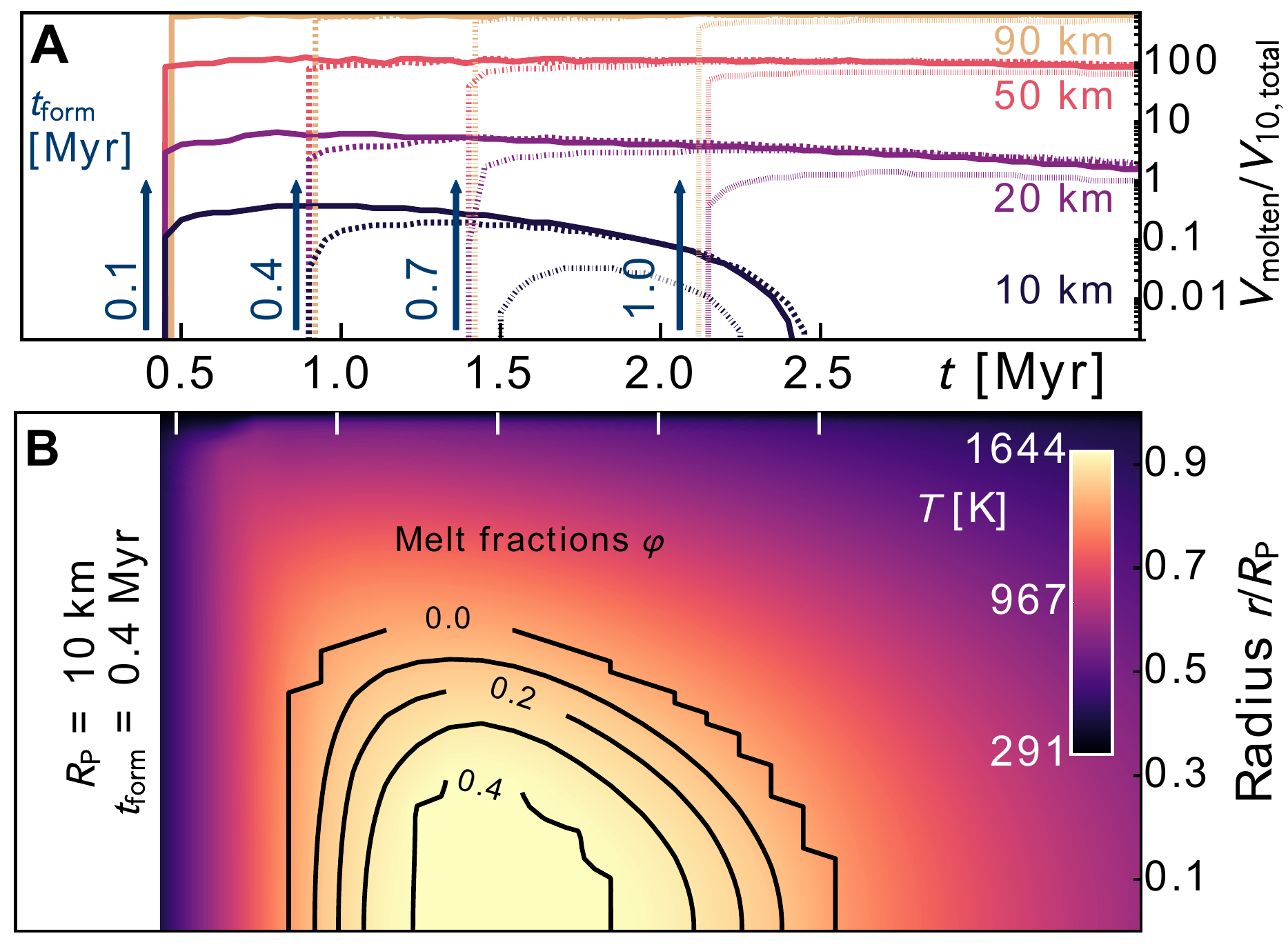}
    \caption{{\bf (A)} Partially molten volume inside planetesimals over time, normalized by the total volume of a body with \RP = 10 km. The low-end mass tail of the planetesimals exhibited partially molten states only during a short time interval, e.g., from 0.5--2.5 Myr (\RP = 10 km). Arrows indicate formation times of associated lines. \RP = 10 km bodies did not exhibit any melt for \tform $\geq$ 1.2 Myr. {\bf (B)} Depth-dependent temperature structure for a planetesimal with \RP = 10 km and \tform = 0.4 Myr. Partial melt fractions (isolines) were sustained for a time period of $\Delta t \sim$ 1.7 Myr after the initial heat-up phase.}
    \label{fig:evolution1}
\end{figure} 

\begin{figure}[tbh]
    \centering
    \includegraphics[width=0.49\textwidth]{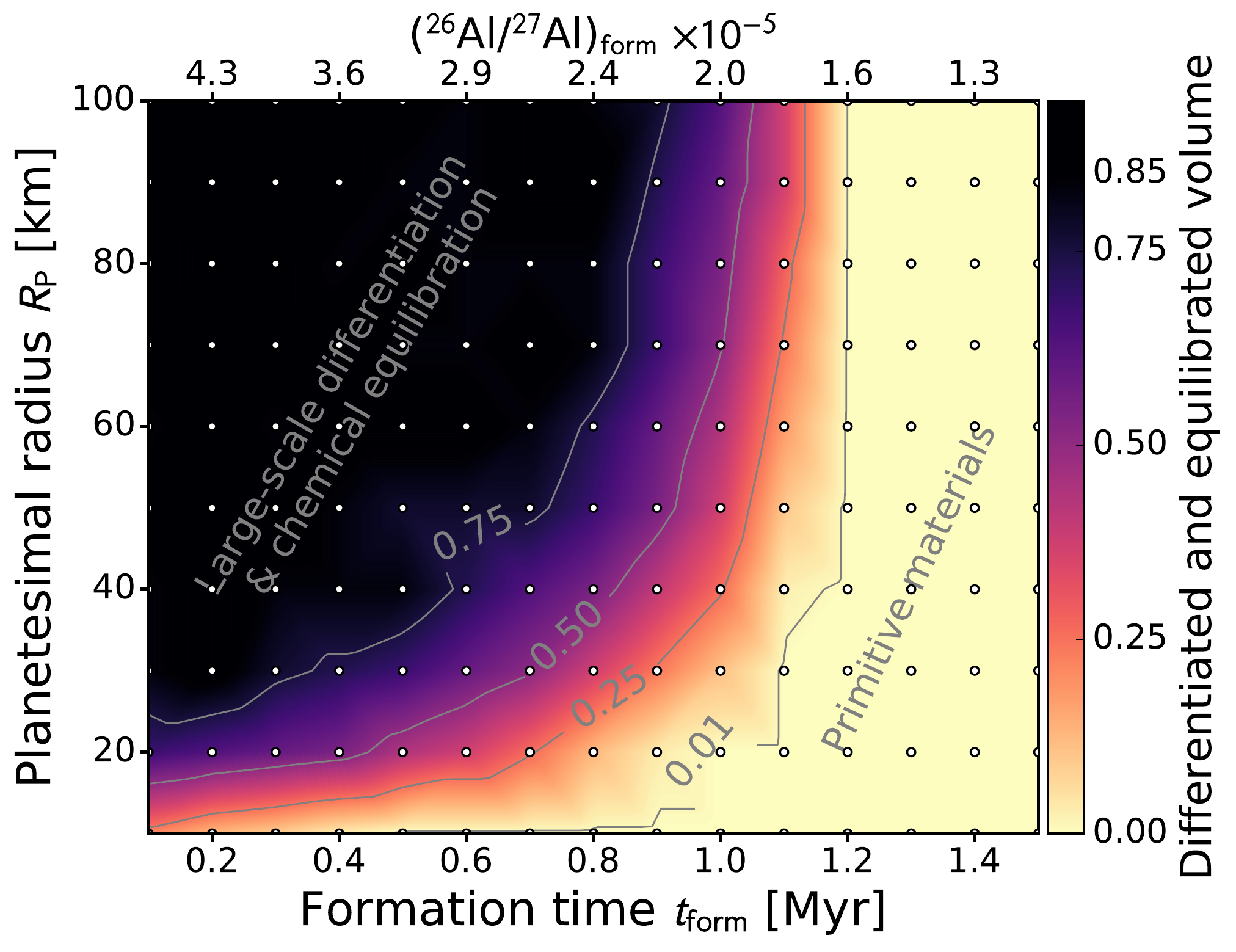}
    \caption{Volume of planetesimals which exceeded melt fractions of $\varphi \geq 0.4$ during the time interval $t$ = 0.1--5 Myr after CAIs, normalized by the total volume of each body. The value at each dot represents the maximum volume fraction throughout a single numerical simulation with the indicated \RP-\tform combination. Especially planetesimals with \RP $>$ 30 km and \tform $\leq 0.9$ Myr after CAIs underwent large scale magma ocean periods and are therefore not eligible as chondrule precursor bodies. For reference, the \alr ratio incorporated into a body at its formation time, for a disk-wide homogeneous \al distribution, is shown at the top.}
    \label{fig:2}
\end{figure}

Figures \ref{fig:evolution1} and \ref{fig:2} show the time-dependent thermal structure of planetesimals due to their interior evolution from \al heating. Figure \ref{fig:evolution1} shows the thermal evolution of one single model with internal melt fractions and in comparison the amount of melt produced within bodies of different sizes and formation times normalized to their total volume. In general, earlier formed and bigger planetesimals exhibited larger heating to cooling ratios, because the radiogenic heat source $^{26}$Al decayed with $t_{1/2}$ = 0.72 Myr, and the surface-to-volume ratios shrank drastically with increasing size of the body. This means that the thermal evolution of the low-mass planetesimals was intrinsically time- and size-dependent. Importantly, a transient regime of silicate material with temperatures around the solidus (\Tsol = 1416 K) within planetesimals existed, which varied drastically with time and depth inside the bodies depending on the planetesimal sizes and formation times. Furthermore, low-mass bodies with radii \RP $\sim$ 10 km exhibited partially molten states only during a narrow time interval $t \sim$ 0.5-2.5 Myr after CAIs. 

Figure \ref{fig:2} shows the maximum fractional volumes of planetesimal models that exceeded the critical melt fraction threshold \phicrit. These planetesimal sub-volumes likely underwent magma ocean stages accompanied by rapid metal-silicate separation and chemical equilibration, as described in Figures \ref{fig:1} and \ref{fig:S1}. In particular, early formed massive planetesimals above $\geq$ 30 km radius with \tform $\leq 0.9$ Myr after CAIs were intensely heated and major parts of their total volume experienced pervasive melting periods. There is, however, a large transition regime with planetesimals mostly experiencing partial melting throughout their interiors (the transition band from black to bright yellow in Figure \ref{fig:2}). Planetesimals that formed later than $ \sim 1.0$ Myr, or alternatively formed with \alr $\leq 1.8 \cdot 10^{-5}$, experienced only minimal melting episodes within the innermost parts of their interior. 

\begin{figure}[tbh]
    \centering
    \includegraphics[width=0.49\textwidth]{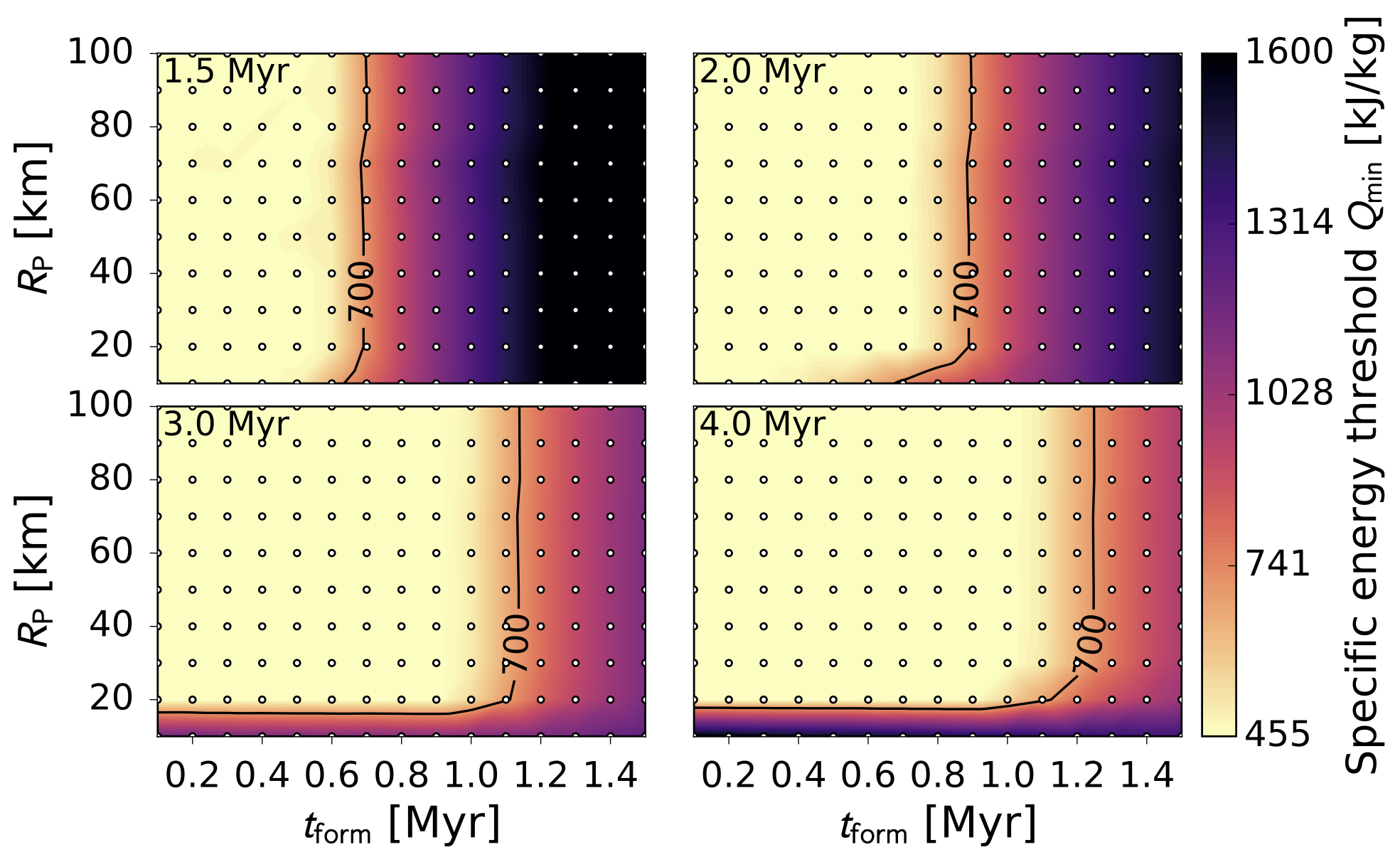}
    \caption{Evolution of the specific energy required at a given time $t$ to raise the temperature in parts of the collisonal debris to above the chondrule formation temperature: $T_{\mathrm{post}} \geq$ \Tchondrule = 1900 K. For this calculation, we assumed homogeneous energy injection across the entire target body. Importantly, the smallest bodies were sufficiently heated during the time interval $t \sim$ 1.5--3 Myr to require impact energies of $Q_{\mathrm{min}} \sim$ 450 kJ per unit mass, after which they cooled down. The evolution of the 700 kJ/kg isocontour shows that until $\sim$ 2 Myr after CAIs small bodies required energies below this value, after 3 Myr after CAIs they required much higher energies to form chondrules.}
    \label{fig:evolution3}
\end{figure}

As a transition to the coupled evolution-collision scenario in Section \ref{sec:results2}, Figure \ref{fig:evolution3} parameterizes the required impact energy for a collision in the super-catastrophic limit. In an idealized scenario, the injected energy must be sufficient to, first heat at least parts of the target body to the required temperatures and, second, disrupt most of the target body into small pieces. In this idealized view, the most energetically favorable source of material for producing chondrules was the center of the target body, as it was hottest due to pre-heating from $^{26}$Al.  

Together with the melt fraction threshold for metal rain-out (\phicrit), this constrains the minimum required impact energy from a two-body encounter to produce chondrules in the collision (Figure \ref{fig:evolution3}). To give a simple example, under the assumption of perfect disruption and energy transfer between impactor and target, the minimal velocity necessary to achieve the critical chondrule formation temperature in the center of the target body is
\begin{eqnarray*}
\Delta v & \geq \sqrt{ \frac{2 M_{\mathrm{tot}}}{m_{\mathrm{imp}}} \cdot \left( \Delta \varphi L\rmu{Si}  + \Delta T c\rmu{p} - \frac{3 G M\rmu{tot}}{5 R\rmu{tot}} \right) },
\end{eqnarray*}
with impactor mass $m\rmu{imp}$, deviations from the required chondrule melt fraction $\varphi\rmu{chondrule} \sim$ 0.86 and temperature \Tchondrule = 1900 K, $\Delta \varphi = \varphi\rmu{chondrule} - \varphi$ and $\Delta T = T\rmu{chondrule} - T\rmu{center}$, latent heat of silicate melting $L\rmu{Si}$ = 400 kJ/kg, pre-collisional material temperature at the target center \Tcenter, silicate heat capacity $c\rmu{p}$ = 1000 J/(kg K), Newton's constant $G$, combined target-impactor mass $M\rmu{tot}$ and combined target-impactor radius $R\rmu{tot}$. As an example, for the case of a super-catastrophic collision of two equally sized planetesimals with \RP = 10 km and internal silicate temperature of \Tcenter = 1600 K, which translates to $\varphi = 1 - (T\rmu{center} - T\rmu{sol})/(T\rmu{liq} - T\rmu{sol}) \sim 0.33$, it would have required impact speeds of $\Delta v \sim$ 1.5 km/s to produce post-impact material with $T\rmu{post} \geq T\rmu{chondrule}$. With one of the two objects being more massive, the injected energy would have increased and thus lower impact speeds would have been sufficient to produce chondrule melt sprays. Still, this calculation and the required energies from Figure \ref{fig:evolution3} define approximate estimates for the impact velocities required in our analytical expression in order to form chondrules from disruptive impacts. This yields roughly $\Delta v \sim$ 1 km/s and is thus presumably higher than the mutual velocities expected to arise from self-stirring of a small planetesimal swarm with radii of up to several tens of kilometers. 

\begin{figure}[!ht]
	\centering
	\includegraphics[width=0.47\textwidth]{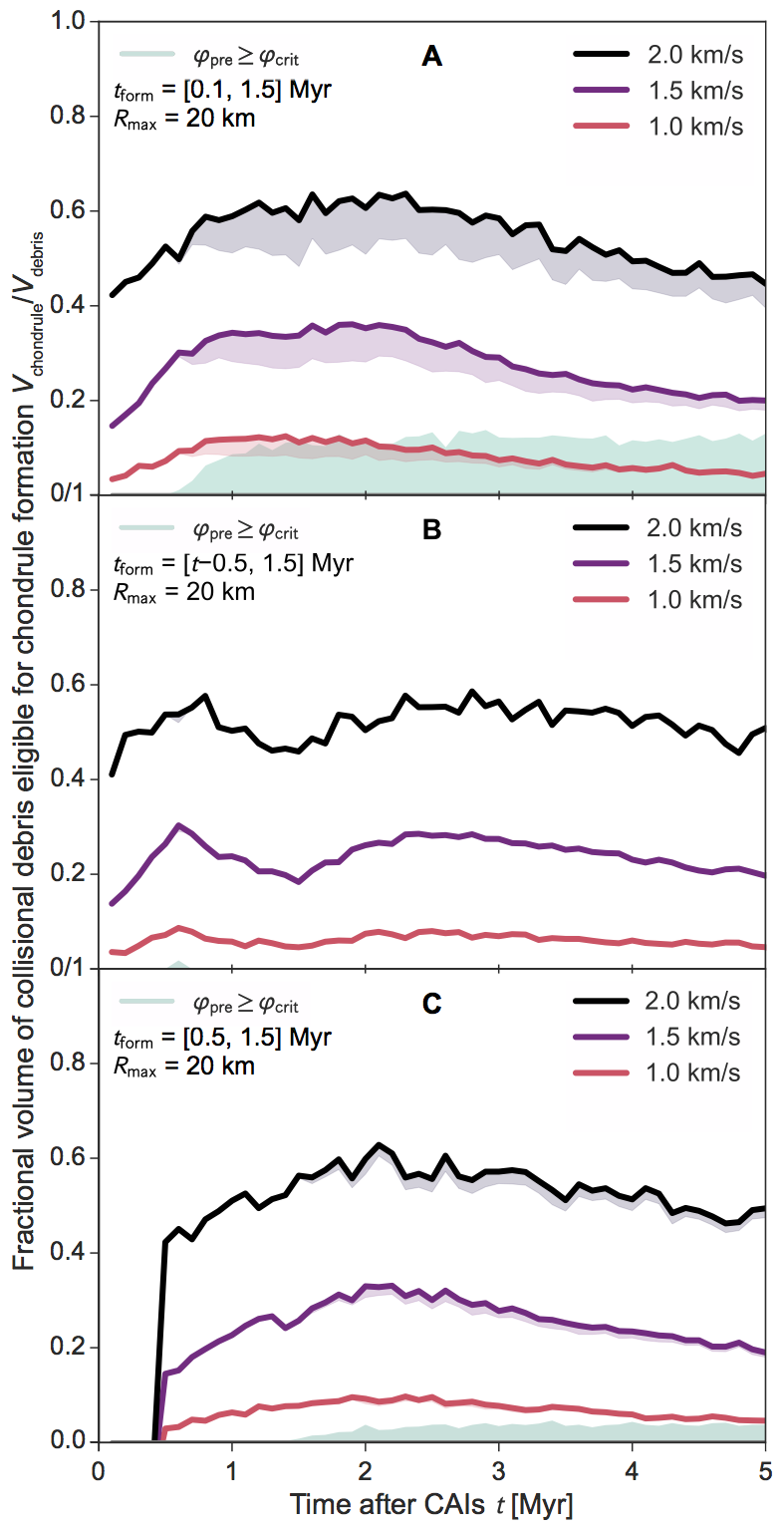}
    \caption{Output of chondrule-eligible collisional debris ($V\rmu{chondrule}$) with post-collision temperatures $T\rmu{post} \geq T\rmu{chondrule} = 1900$ K over time from randomized super-catastrophic collisions, normalized to the total volume of generated debris ($V\rmu{debris}$). Each line in each subplot corresponds to a single Monte Carlo simulation with varying parameters. Black lines are runs with impact velocities $\Delta v = $ 2.0 km/s, purple lines $\Delta v = $ 1.5 km/s and red lines $\Delta v = $ 1.0 km/s. Shaded areas below these lines indicate the variation from using a different treshold for metal rain-out/chemical equilibration (\phicrit), with the lower bound \phicrit = 0.4 and the upper bound \phicrit = 0.5. Here, we consider bodies with \RP = 10--20 km and {\bf(A)} \tform = 0.1--1.5 Myr, {\bf(B)} \tform randomly drawn from the time interval $\Delta t$ = 0.5 Myr before the collision time, and {\bf(C)} \tform = 0.5--1.5 Myr. Green shaded areas show metal-depleted debris, that means, material originating from source regions with $\varphi \geq$ \phicrit before the collision. The amount of metal-depleted material decreased significantly if collisional processing was efficient and the average dwell time of intact planetesimals was short {\bf(B)}, or when planetesimal formation was suppressed during the early disk phase ({\bf C}, \nameref{sec:suppl}).}
    \label{fig:collision1}
\end{figure}

\subsection{Collisional processing}
\label{sec:results2}

To explore the effects of varying impact speeds on the planetesimal population, we developed a Monte Carlo approach to model the time-dependent influence of increased internal energies in the parent bodies on potential chondrule material in the collision aftermath (Section \ref{sec:montecarlo}). The results from several simulation runs for planetesimal swarms with \RP = 10--20 km are shown in Figure \ref{fig:collision1}. In general, higher collision velocities increase the output of melt from the collision. For several cases (like in Figure \ref{fig:collision1}A), the thermal evolution from $^{26}$Al heating produced a peak in eligible chondrule material output for constant collision velocities at around $t \sim$ 2 Myr. If planetesimals were allowed to form during the whole interval from \tform = 0.1--1.5 Myr (Figure \ref{fig:collision1}A), the output of metal-free material became noticeable after $t \geq$ 0.7 Myr. If the colliding families were preferentially formed at later time intervals, metal-free output became insignificant or virtually non-existent (Figure \ref{fig:collision1}B,C and \nameref{sec:suppl}). For bigger planetesimal regimes with radii up to 30, 50 or 100 km (\nameref{sec:suppl}) this trend holds, while the output of chondrule forming material relative to material output with $T\rmu{post} <$ 1900 K decreased.

\begin{figure*}[htb!]
	\centering
	\includegraphics[width=0.89\textwidth]{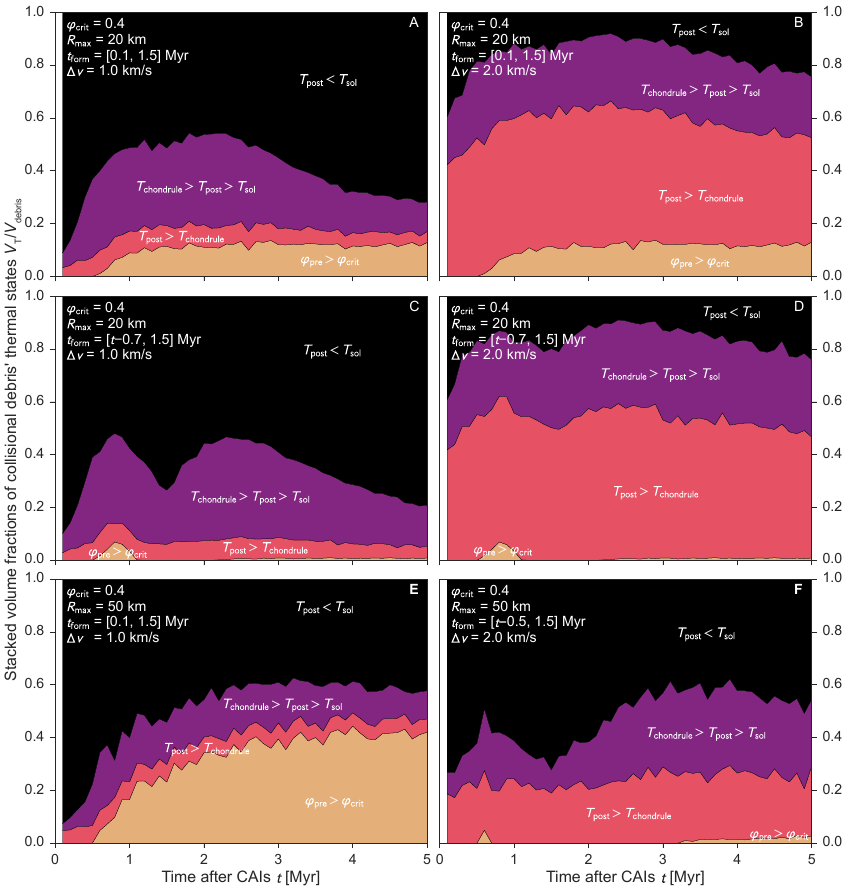}
    \caption{Thermal state of all post-collisional debris material over time for planetesimal swarms with \RP = 10--20 km {\bf(A-D)} and \RP = 10--50 km {\bf(E,F)}, metal preservation threshold criterion \phicrit = 0.4 and impact velocities of $\Delta v =$ 1 km/s {\bf(A,C,E)} and $\Delta v =$ 2 km/s {\bf(B,D,F)}. Each subplot corresponds to a single Monte Carlo simulation. Different colors represent post-collisional material volume with varying thermal state, normalized to the volume of the total debris generated during continuous collisional recycling. The thermal distribution of the collision aftermath depended on the specific conditions during the impact, which can change the ratio of unmelted (black, \Tpost $<$ \Tsol) to melted (purple, \Tchondrule $>$ \Tpost $>$ \Tsol) debris and the output of chondrules (red, \Tpost $>$ \Tchondrule). Metal-depleted material (due to metal-silicate segregation prior to collision) is depicted in yellow ($\varphi\rmu{pre}$ $>$ \phicrit). {\bf(B,D,F)} all show eligible-chondrule material of $V\rmu{chondrule}/V\rmu{debris} =$ 0.2--0.5\% throughout the time evolution. Subfigures {\bf(C,D,F)} demonstrate the influence of a variable formation time interval in the model, representing an effective collisional grinding such that no bodies formed earlier than 0.7 Myr {\bf(C,D)} or 0.5 Myr {\bf(F)} before the collision time participate in the cycle. In the setting shown here, this generates some differentiated debris at $\sim$ 0.9 Myr {\bf(C,D,F)}, but for the rest of the disk-phase prevented overheated material entering the debris cycles. {\bf(E)} shows that larger planetesimal families readily reached debris states with a high fraction of differentiated debris (and potentially basaltic spherules as collision output). Importantly, we do not take the outcome from previous collision cycles into account, treating each time step as independent from the ones before. That implies that material labeled `red' at $t =$ 1.0 Myr can become `black' in the next time step and vice versa for all label mutations. See \nameref{sec:suppl} for further parameter variations.}
    \label{fig:thermal1}
\end{figure*} 

It is important to note that the collisional debris showed a broad thermal distribution. Material from the inner parts of the body was heated to higher temperatures and thus higher melt fractions, whereas the outer parts of the colliding bodies may have remained cool and resulted in unmelted debris, which is not seen in chondritic meteorites. To address this issue, we have quantified the post-collisional thermal distribution for several parameter combinations and thus different collision families in Figure \ref{fig:thermal1} (more in \nameref{sec:suppl}). In order to `suppress' the generation of a vast amount of differentiated material \citep[which yields basaltic droplets unlike the chondrules observed in the meteoritic record,][]{asphaug2017} the make-up and dynamics of the colliding planetesimal swarm must be either (i) primarily composed of low-mass planetesimals, (ii) formed late, or (iii) feature a low dwell time before collisional recycling. The latter would correspond to a high encounter probability, which cools the material via total disruption, fragmentation or partial break-up \citep{2013M&PS...48.2559C}.

The thermal distribution, and thus the ratios of melted, partially melted and unmelted debris, crucially depends on the localization of energy transfer during the collision. In our super-catastrophic model, however, in which the energy is distributed across the entire body, the pre-collision temperature solely determines the temperature deviations of the post-collision debris. Depending on the formation times, sizes and recycling efficiencies of the colliding planetesimals, the ratio of melted to unmelted debris may shift further (\nameref{sec:suppl}).

%%%%%%%%%%%%%%%%%%%%%%%%%%%%%%%%%%%%%%%%%%%%%%%%%%%%%%%%%%%%%%%%%%%%%%%%%%%%%%%%%%%%%%%%%%%%%%
%%%%%%%%%%%%%%%%%%%%%%%%%%%%%%%%%%%%%%%%%%%%%%%%%%%%%%%%%%%%%%%%%%%%%%%%%%%%%%%%%%%%%%%%%%%%%%
%%%%%%%%%%%%%%%%%%%%%%%%%%%%%%%%%%%%%%%%%%%%%%%%%%%%%%%%%%%%%%%%%%%%%%%%%%%%%%%%%%%%%%%%%%%%%%
%%%%%%%%%%%%%%%%%%%%%%%%%%%%%%%%%%%%%%%%%%%%%%%%%%%%%%%%%%%%%%%%%%%%%%%%%%%%%%%%%%%%%%%%%%%%%%
%%%%%%%%%%%%%%%%%%%%%%%%%%%%%%%%%%%%%%%%%%%%%%%%%%%%%%%%%%%%%%%%%%%%%%%%%%%%%%%%%%%%%%%%%%%%%%
%%%%%%%%%%%%%%%%%%%%%%%%%%%%%%%%%%%%% DISCUSSION %%%%%%%%%%%%%%%%%%%%%%%%%%%%%%%%%%%%%%%%%%%%%
%%%%%%%%%%%%%%%%%%%%%%%%%%%%%%%%%%%%%%%%%%%%%%%%%%%%%%%%%%%%%%%%%%%%%%%%%%%%%%%%%%%%%%%%%%%%%%
%%%%%%%%%%%%%%%%%%%%%%%%%%%%%%%%%%%%%%%%%%%%%%%%%%%%%%%%%%%%%%%%%%%%%%%%%%%%%%%%%%%%%%%%%%%%%%
%%%%%%%%%%%%%%%%%%%%%%%%%%%%%%%%%%%%%%%%%%%%%%%%%%%%%%%%%%%%%%%%%%%%%%%%%%%%%%%%%%%%%%%%%%%%%%
%%%%%%%%%%%%%%%%%%%%%%%%%%%%%%%%%%%%%%%%%%%%%%%%%%%%%%%%%%%%%%%%%%%%%%%%%%%%%%%%%%%%%%%%%%%%%%
%%%%%%%%%%%%%%%%%%%%%%%%%%%%%%%%%%%%%%%%%%%%%%%%%%%%%%%%%%%%%%%%%%%%%%%%%%%%%%%%%%%%%%%%%%%%%%
\section{Discussion} 
\label{sec:discussion}

\subsection{Constraints from the interior evolution}
\label{sec:interior}

The scaling analysis performed in Section \ref{sec:results1} (Figures \ref{fig:1} and \ref{fig:S1}) demonstrates that high melt fraction regions in radiogenically heated planetesimals rapidly evolve to physically differentiated and chemically equilibrated states. We note that the scalings used in this section are based on thermal driving forces and neglect potential other effects like magnetic fields \citep{2015aste.book..533S} or rotation \citep{2009Natur457301K,2015JGRB1207508M} that may alter the regimes. However, any potential changes to the convection regime within magma ocean planetesimals have two effects. If turbulence is less vigorous than derived here, Fe,Ni metal settling would be even more rapid, since the energy to suspend these droplets would decrease. In the opposite case, turbulence would be more vigorous and accelerates chemical equilibration, because the diffusion time scale derived from the Kolmogorov microscales diminishes. Therefore, we conclude that any planetesimals that can serve as eligible precursor bodies for chondrule formation in a collision event cannot have been fully-molten to above the rheological transition throughout a large fraction of their interior.

In general, it is important to note that the thermal evolution -- and thus degree of differentiation and chemical homogenization -- forms a continuum \citep[Figures \ref{fig:evolution1}, \ref{fig:2} and \ref{fig:evolution3}; compare][]{lichtenberg16a}. Therefore, the number of bodies that experienced substantial radiogenic preheating depends on the \emph{local} planetesimal size frequency distribution, formation rate and recycling efficiency over time, in particular during the first 2 Myr after CAIs, during which the radiogenic heating from $^{26}$Al was the predominant contributor to the internal evolution. From a thermomechanical point-of-view, the low-mass tail of planetesimals or bodies formed at sub-canonical \al abundances (the transition region in Figure \ref{fig:2} and bodies labeled with `primitive materials') barely incorporated enough $^{26}$Al to reach temperatures near the solidus throughout most of their volume, and presumably never reached the rheological transition at melt fractions $\varphi_{\mathrm{crit}} \approx $ 0.4--0.6 \citep{2009GGG....10.3010C} to develop an internal magma ocean \citep{lichtenberg16a}. 

Regarding the Fe,Ni metal abundances in and around chondrules, it is important to note that recent laboratory experiments demonstrated the trapping of metallic liquids in planetesimal mantles with low silicate melt fractions \citep{2009E&PSL.288...84B,2011GGG....12.3014R,holzheid2013sulphide,2015E&PSL.417...67C,2016AmMin.101.1996T}. They showed that, first, in the regime below the silicate solidus, the high interfacial energy and wetting angle between metal-sulfide melts and solid silicate mantle minerals preclude efficient metallic core formation \citep{2009E&PSL.288...84B,2011GGG....12.3014R}. Second, it was shown that in the regime of modest silicate melt fractions, mobile basaltic melts reduce the interconnectivity and segregation of metal-sulfide liquids under deformation conditions with varying strain rates. This leaves some metallic liquid stranded in the olivine matrix until the rheological transition is reached \citep{holzheid2013sulphide,2015E&PSL.417...67C,2016AmMin.101.1996T}. Above this threshold the silicate viscosity drops by orders of magnitudes and metal-silicate differentiation by gravitational settling becomes efficient \citep[Figure \ref{fig:1} and][]{2012AREPS..40..113E}. From these experiments, we conclude that complete metal-silicate segregation in planetesimals required significant melt fractions, likely around and higher than the rheological transition. Therefore, small (\RP $\sim$ 10--30 km) and/or late formed (\tform $\geq$ 0.7--1.0 Myr after CAIs) planetesimals, with their presumably low melt fractions and thus incompletely differentiated interiors, can retain substantial metal abundances and chemical and isotopic heterogeneities distributed throughout most of the planetesimal volume. This qualifies these planetesimals as eligible chondrule precursor material.

\begin{figure*}[tbh]
    \centering
    \includegraphics[width=0.69\textwidth]{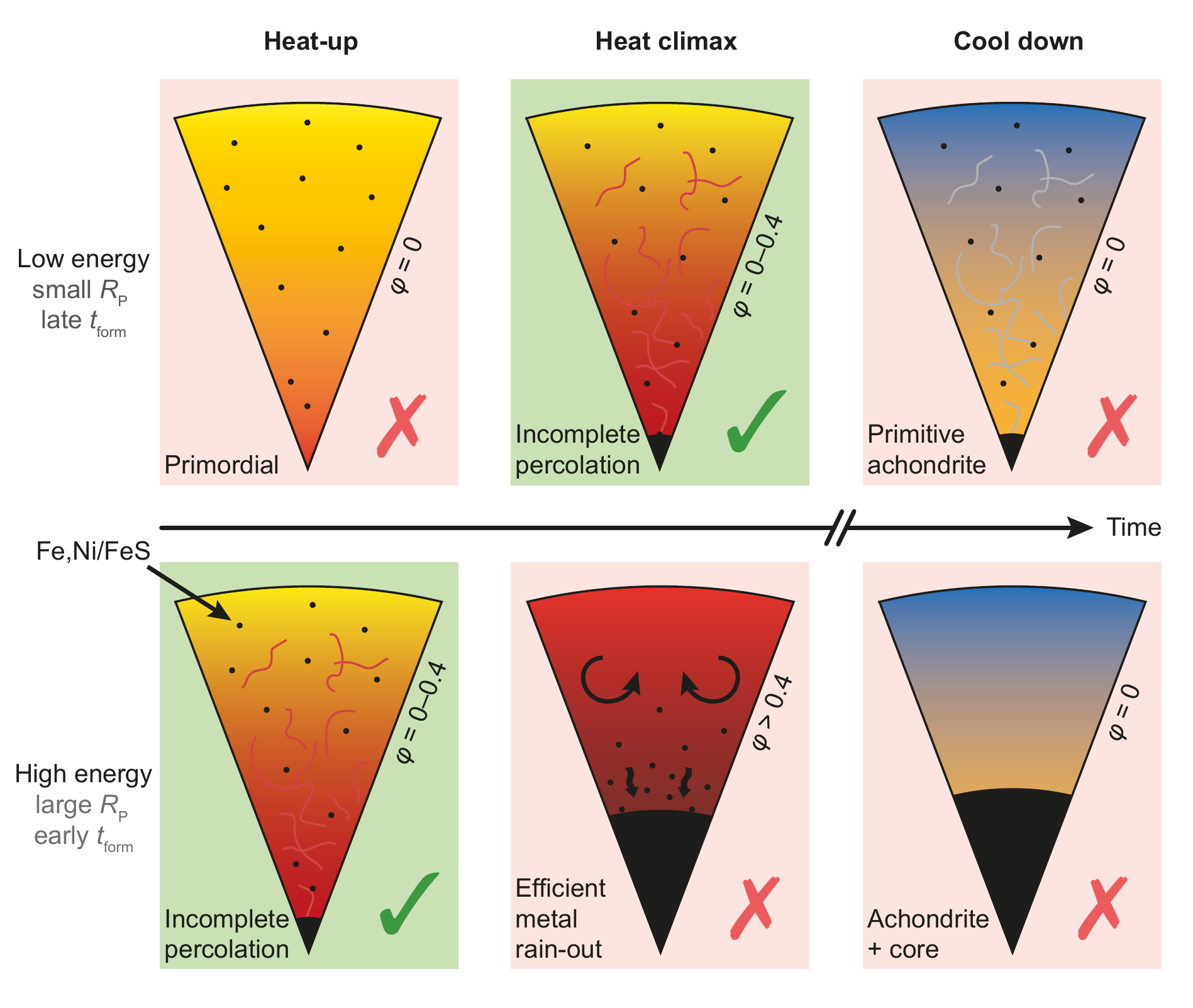}
    \caption{Schematic illustration of the qualitative thermomechanical planetesimal evolution regimes. {\bf(Top)} Low-energy bodies with relatively small radii or late formation times, which were eligible chondrule precursor bodies (highlighted in green) around their heat climax (compare Figure \ref{fig:evolution1} and Figure \ref{fig:evolution3}). {\bf(Bottom)} High-energy bodies with either large radii or early formation times, which were eligible precursor bodies only during their brief initial heat-up phase. Models highlighted in red either do not feature high-enough radiogenic pre-heating (and thus would have needed implausible high impact velocities) or have lost their primordial metal abundances due to efficient metal-silicate segregation processes.}
    \label{fig:illustration1}
\end{figure*}

Efficient collisional recycling of small or late-formed planetesimals is needed to produce large quantities of metal-bearing and chemically heterogeneous chondrules to populate the asteroid main belt with chondrite parent bodies. For that, planetesimals with modest internal melt fractions must have been abundant in the early solar system. An additional source of chondrules may have come from larger planetesimals that experienced impacts in narrow time windows during their heat-up phase, when the radiogenic pre-heating was sufficient but before reaching the magma ocean phase. We have qualitatively summarized the thermomechanical planetesimal regime that may be capable of chondrule formation in the aftermath of a collision in Figure \ref{fig:illustration1}. The regime we propose as potential chondrule precursor bodies is highlighted in green. In general, small bodies or bodies with sub-canonical \al abundances were more likely to be heated to suitable temperatures at around their thermal maxima and to not reach melt fractions above the rheological transition throughout most of their interiors. In comparison, larger bodies featured chondrule-eligible interior states only during their initial heat-up phases.

\subsection{Accretion and dynamical recycling}
\label{sec:accretion}

The results from the evolution-collision model (Section \ref{sec:results2}, Figures \ref{fig:collision1}, \ref{fig:thermal1} and \nameref{sec:suppl}) underline two points. First, planetesimals preheated by the decay of \al require less energetic collisions than non-preheated bodies to produce chondrules in the collisional aftermath. If the bodies participating in the collision have not reached the magma ocean phase, droplets resulting from the impact can satisfy the constraints from chondrule textures, i.e., subsonic impact velocities in order to avoid shock textures in the resulting material \citep{asphaug2017}. Therefore, we require velocities higher than the two-body escape speed but lower than for the case of cold planetesimals. Such velocities may be achieved during the gas disk phase (see further down). Second, Figure \ref{fig:thermal1} and the additional figures for variable parameter regimes (\nameref{sec:suppl}) demonstrate that the outcome of any collisional regime can be highly variable, depending on the local size frequency distribution (SFD), the internal state of the planetesimals and the dynamical regime of the swarm participating in the collisional processing. It underlines the necessity of simultaeneously considering the {\em global} dynamical and {\em local} formation, growth and destruction mechanisms in astrophysical models of planet formation in the early solar system, which we therefore discuss here.

The first planetesimals likely formed according to a specific size frequency distribution \citep{2007Natur.448.1022J,2008ApJ...687.1432C,2010Icar..208..505C,2016ApJ...822...55S}. Recent planetesimal formation models can produce bodies via the streaming instability or turbulent concentration mechanisms within the first few million years in the solar nebula \citep{2008ApJ...687.1432C,2010Icar..208..505C,2015A&A...579A..43C,2016A&A...594A.105D,2017A&A...602A..21S}. Latest estimates of the initial size frequency distribution of planetesimals from the streaming instability mechanism \citep{2015SciA....115109J,2016ApJ...822...55S,2017arXiv170503889S} converge on a power law $\dmath N / \dmath \RPmath \sim \RPmath^{-q}$, with the number of bodies $N$, the planetesimal radius \RP, index $q = 2.8$ and without an obvious lower cut-off. These estimates are consistent with the current distribution in the asteroid belt if accretional growth of small bodies via collisions and/or pebble accretion is considered \citep{2011Icar..214..671W,2014ApJ78022L,2015SciA....115109J,2017Icar281459M}. In such a birth-SFD, the bulk of the mass resides in massive planetesimals, while the absolute number of low-mass planetesimals exceeds the number of massive bodies by orders of magnitude. Therefore, collisions among low-mass members of the SFD outnumber interactions with one or between two massive members, even if gravitational focusing is considered. This low-mass planetesimal collision regime was most vulnerable to disruption during collisions and thus presumably created much debris from hit-and-run, erosive or (super-)catastrophic interactions \citep{2010ChEG70199A,2012ApJ74579L}.

The collisional dynamics and, therefore, the prevailing impact parameters crucially depend on the ambient disk conditions and the nature of planet(esimal) growth \citep{1989Icar77330W,1993Icar..106..190W,1996Icar123180K,1998Icar..131..171K,2011Icar..214..671W,2015SciA....115109J}. Importantly, in a growing planetesimal swarm the parameter dominating the mean impact velocity among bodies is the size of the largest body. The largest body stirs the velocity dispersion in the swarm to its own escape velocity and the smallest bodies reach the highest relative velocities \citep[e.g.,][]{2011ApJ...728...68S}. During the disk stage, the velocity dispersion can be highly reduced due to gas damping, which complicates reaching sufficient impact velocities to generate chondrules melts. This problem originally motivated the idea of chondrule formation from fully-molten planetesimals and the sole reliance on \al as a heat source for chondrule formation \citep{2011E&PSL.308..369A,2012M&PS...47.2170S}.

Therefore, producing chondrites as $N$th generation planetesimals from the collisional recycling of radiogenically preheated but undifferentiated planetesimals required, first, continuous collisional reprocessing during the first few million years after the formation of the Sun. Second, sufficient velocity dispersions above the mean two-body escape velocity of the lowest-mass members of the collisional planetesimal swarm ($v\rmu{esc} \lesssim$ 0.1 km/s in the size regime evaluated here) must have been triggered. Potential stirring mechanisms to enable these enhanced mutual encounter velocities for low-mass planetesimals are manifold, for instance early formation of planetary embryos like Mars \citep{2011Natur.473..489D,2016ApJ8168H}, migration of giant planets \citep{2011Natur.475..206W,2016ApJ...833...40I} and/or giant planets' forming cores \citep{2016MNRAS.458.2962R}, resonant excitations \citep{1998Sci...279..681W} or implantation of planetesimals via scattering into the main belt region \citep{2006Natur.439..821B}. These mechanisms depend strongly on the ambient gas density and become more efficient as the solar nebula disperses over time, leading to decreased gas damping and allowing for higher mutual velocities.

The absolute volume of low-mass planetesimals in our solar system was presumably minor compared to the material within massive bodies \citep{2015SciA....115109J, 2016ApJ...822...55S}. However, it was subject to most destructive collision events among planetesimals in terms of absolute numbers. Due to their larger cross-section and enhanced gravitational focusing, the largest bodies accreted the fastest and thus presumably accumulated to form the terrestrial planets in the inner solar system \citep{1989Icar77330W,1997Icar128429W}. The low-mass bodies were dynamically excited by the larger body-size population, which enhanced encounter rates. Depending on the planetesimal number and how they actually arrived at specific locations in the disk (for instance, in-situ formation versus implantation), the debris from low-mass collisions can dominate the total {\em local} solid density and provide the environment for the make-up of chondrite parent bodies as a result of collisionally recycled low-mass or sub-canonical-\al planetesimals. 

Recent observations and theoretical considerations estimate that the solid pile-up within 'sweet spots' in inner disk regions facilitated planetesimal formation in confined bands \citep{2016A&A...594A.105D,2016ApJ...820L..40A,2016PhRvL.117y1101I,2017A&A...602A..21S,2017ApJ...839...16C}. In such narrow planetesimal birth regions with high solid density, efficient collisional processing can be enhanced because of higher collisional cross-sections and average encounter rates compared to the classical picture of disk-wide planetesimal formation. In our model, collisions of chondrule-eligible planetesimals preferentially produced a chondrule formation peak at around $t \sim$ 2 Myr after CAIs (Figure \ref{fig:collision1} and \nameref{sec:suppl}). Although the exact timing depends on the parameter choice, this peak is in very good agreement with the radiogenic ages determined for chondrules in meteorites \citep[][see chondrule geochronology Section \ref{sec:geochem}]{2009Sci...325..985V,2014E&PSL.398...90M,2015GMS...212....1C}. In our models, the peak reflects the temperature climax of planetesimals heated from \al decay \citep[as suggested in][]{2012M&PS...47.2170S}. Many chondrules may thus reflect collisional debris of low-mass and preheated planetesimals, whose material was not swept up by early oligarchs and survived in smaller bodies that comprise the asteroid belt today. 

The asteroid belt today is significantly depleted in mass relative to the terrestrial and giant planet regions of the solar system. Therefore, either the region was dynamically depleted early-on or the mass depletion must be primordial \citep{2006Natur.439..821B,2011Natur.475..206W,2016ApJ...833...40I,2016A&A...594A.105D,2016JGRE..121.1962M}. In the latter case, this may be a suitable environment for the kind of dynamical processing we propose here and the complex transition from S- to C-type asteroids \citep{2015aste.book...13D}. For instance, recent studies suggest early mixing of silicate and ice-rich planetesimals \citep{2016SciA....2E1001M}. In this picture, planetesimals may either form in lower numbers in the asteroid belt region or can be implanted from inner and outer disk regions. Thus, they would originate from distinct source reservoirs, as was suggested for iron meteorites which may have formed pre-dominantly in the inner disk region \citep{2006Natur.439..821B}.

\subsection{Collision physics}
\label{sec:collision_physics}

In general, a complicating issue for estimates of debris generation is that the effective amount of material excavated during the collision depends on a variety of impact parameters, such as the impact velocity, impact angle, mass ratio of the colliding bodies and material compositions \citep{2009Icar199542L,2010ChEG70199A,2012ApJ74579L,2016Icar27585M}. In addition to the amount of material ejected during the collision, it is important to understand the energy distribution in the ejected fragments/droplets. This, in turn, determines the thermal histories of chondrules produced in the collision fragments (Figures \ref{fig:thermal1} and \nameref{sec:suppl}) and is crucially dependent on the energy localization during the impact, for which high-resolution three-dimensional numerical models are required. The derivation of scaling laws that account for the combined effects of ejecta size and energy distribution is a long-term goal of the impact modeling community \citep[e.g.,][]{2017Icar..296..239S}. To our knowledge, at present there are no scaling laws that cover a large parameter space and can be used to couple the interior evolution of planetesimals prior to collision with the energy injection and material ejection during the collision.

Finally, we want to point out that, in the context of our semi-analytic evolution-collision model impacts usually require $\Delta v \gtrsim$ 0.5 km/s to generate chondrule eligible material. However, using more advanced numerical collision models utilizing more realistic energy localization together with a statistical or $N$-body growth model \citep[e.g.,][]{2009Icar..204..558M,2011ApJ...728...68S,2015ApJ...813...72C,2015SciA....115109J,2017Icar281459M} of the planetesimal swarm will enable much lower collision velocities. As demonstrated by \citet{2017ApJ...834..125W}, collisions among unheated planetesimals may already be close to generate chondrule eligible temperatures in the collision aftermath, therefore coupling both energy sources (collision and radiogenic heating) will be even more capable of producing the correct conditions.

\subsection{Geochemical perspective}
\label{sec:geochem}

In this section we review some important geo- and cosmochemical constraints for the origin of chondrules and a potential collisional origin. Many interpretations of the chondrule record with respect to chondrule formation via collisions were already discussed in-depth by \citet{2012M&PS...47.2170S} and partly in \citet{2016JGRE..121.1885C}. Even though these authors focused on fully-molten planetesimals as chondrule precursors, many of their interpretations also apply to the more moderate and realistic scenario of a radially heterogeneous interior evolution of colliding planetesimals that did not reach the magma ocean stage. Therefore, we will not repeat these arguments here, but instead focus on issues that emerged in recent years or are of direct consequence for our plead toward a more nuanced debate of collision models for chondrule formation.

\subsubsection*{Chondrule geochronology}

Based on various radiometric dating techniques, chondrules formed during the first $\sim$ 4--6 Myr after CAIs \citep{2014mcp..book...65S,2015GMS...212....1C}. The exact details, however, are debated. There is an on-going debate in the cosmochemical community as to whether \al was heterogeneously distributed in the protoplanetary disk \citep{2011ApJ...735L..37L,2015E&PSL.420...45S,2016PNAS..113.2011V,kleine2017}. The consequences of a heterogeneous distribution are far reaching. In the following we discuss the implications for chondrule formation by collisions of preheated planetesimals for the case of (i) a homogeneous distribution, and (ii) a heterogeneous distribution. 

\begin{enumerate}[(i)]

	\item A homogeneous \al distribution entails that precise Al-Mg ages of chondrules can be obtained and these indicate a time gap of $\Delta t$ = 0.5--1.0 Myr between the formation of CAIs (at $t \sim 0$ Myr) and the onset of chondrule formation \citep{2009Sci...325..985V,2012M&PS...47.1108K,nagashima201426al,2015GMS...212....1C,2015GeCoA.160..277V}. In Section \ref{sec:accretion} we argued that the apparent peak in chondrule formation ages may be linked to the interior heat climax of \al-heated planetesimals. However, for some parameter combinations our models do not produce a gap during the first Myr after CAI formation. The apparent time gap between CAIs and chondrules could therefore reflect protracted planetesimal formation after CAIs (Figure \ref{fig:2}) and thus reduced \al inventories or alternatively insufficient debris ejection from collisions during the early phase of high gas-damping. Delayed planetesimal formation may be due to increasing dust-to-gas ratios with time due to photoevaporation \citep{2009ApJ...704L..75J,2017ApJ...839...16C}. Decreasing ambient gas densities also allow for higher mutual velocities.

	\item Based on a heterogeneous distribution of \al in our solar system, the cosmochemical record provides evidence for an extended period of chondrule formation, starting contemporaneously with CAI formation over 3--4 Myr \citep{2011ApJ...735L..37L,2012Sci...338..651C,2015E&PSL.420...45S,2016PNAS..113.2011V,2017GeCoA.201..345C}. These results are inferred from Pb-Pb ages of individual chondrules. In this context, no gap needs to be reproduced, but the \al inventory may be sub-canonical everywhere in the disk except the CAI forming region. This would shift the thermomechanical regimes in Figure \ref{fig:2} but would still allow for substantial radiogenic preheating depending on the local inventory of \al at the time of planetesimal formation or reaccretion.

\end{enumerate}

\subsubsection*{Nucleosynthetic, chemical and petrographic constraints} 

Chemical and isotopic complementarity is a concept based on elemental and isotopic studies of chondrules and matrix. Various elemental and isotope compositions may be distinct in matrix and chondrules of a specific chondrite, but, when mixed together, complement each other to nearly CI-like composition \citep{2005PNAS..10213755B,2008E&PSL.265..716H,2010E&PSL.294...85H,2015E&PSL.411...11P,2016GeCoA.172..322E}. This extends to W and Mo isotope variations derived from presolar carriers, which may also be complementary in `matrix' (defined as fine-grained dust between chondrules) and chondrules \citep{2015E&PSL.432..472B,2016E&PSL.454..293B,2016PNAS..113.2886B}. These relations are often claimed to rule out particular chondrule formation mechanisms, such as a collisional origin of chondrules. Indeed, \emph{if} complementarity of matrix and chondrules is real and indicates a genetic heritage linked by the chondrule formation process itself, this provides severe constraints on \emph{every} chondrule formation mechanism suggested to date. The technical details and interpretations of the chondrule-matrix complementarity hypothesis are controversially debated in the community and out of the scope of this paper.

In recent years, it was found that terrestrial bodies in the solar system exhibit distinct nucleosynthetic isotope signatures in, e.g., Zr \citep{2011LPICo1639.9085S,2015GeCoA.165..484A}, Ni \citep{2008E&PSL.272..330R,2012ApJ...758...59S}, Cr \citep{2007ApJ...655.1179T,2009Sci...324..374T,olsen2016magnesium} and Mo \citep{2011E&PSL.312..390B}. How can this be reconciled with the collisional origin of chondrules we put forward in this manuscript? In any accretion scenario the most massive planetesimals and embryos preferentially served as the early precursors of the planets. Therefore, the most massive bodies were unavailable as meteorite parent bodies because they either seeded the planet formation processes themselves or preferentially interacted with the accreting protoplanets due to gravitational focusing and enhanced geometrical encounter rates. Since accretion mechanisms like planetesimal agglomeration or pebble accretion become less efficient for smaller bodies, debris from low-mass planetesimal collisions was presumably insufficient in mass to act as a seed for planet formation. In light of chondrule formation from collisions of low-mass planetesimals, chondrites can then be interpreted as left-over material, which did not end up in planets. Instead, it either formed small parent bodies by itself or was accreted onto other relatively low-mass bodies. This implies that the materials sampled in the meteoritic record were not important contributors to the chemical bulk planet compositions in the solar system. Importantly, in this picture the chondrites sample qualitatively different material than represented in the Earth and the other terrestrial planets. This is consistent with nucleosynthetic signatures identified in meteorites, which are distinct from those of bulk Earth \citep{2011E&PSL.312..390B, 2015GeCoA.165..484A,2016LPI....47.2252P}.

In our model, age differences between chondrules of single chondrites can be attributed to, for instance, the storage of chondrules in outer parts of planetesimals, later liberation during disruption of the body and mixing with newly formed chondrules in the subsequent reaccretion of a new parent body. On the other hand, if age gaps in each chondrite group are narrow \citep[as suggested by][]{2012M&PS...47.1157A} chondrule variability in a parent body can be obtained by mixing of debris ejecta from several chondrule-forming collisions in one annulus. Furthermore, most material eligible for chondrule formation, i.e., collisional debris, which is heated to \Tpost $>$ 1900 K \citep{2016JGRE..121.1885C, 2008Sci...320.1617A, 2015GeCoA.160..277V} during one of the collisional cycles, not necessarily (fully) re-melts during the reprocessing. This allows chondrules to preserve relict grains -- in agreement with the chondrule record \citep{2012M&PS...47.1176J} -- and generates further chondrule diversity due to variable interaction of the ejected fragments with molten material and vapor in the impact plume \citep{2015GeCoA.160..277V}. Moreover, each chondrite parent body sampled a distinct, isolated reservoir without much mixing with those of other chondrite parent bodies \citep{2012M&PS...47.1176J}. In the context of our model, this is a natural consequence if each chondrite parent body sampled a distinct band defined by the pile-up sweet spots for accretion \citep{2016A&A...594A.105D,2017A&A...602A..21S,2017ApJ...839...16C} or implantation \citep{2006Natur.439..821B}. 

In our model, the ejection of material in the aftermath of the collision resulted in disconnected droplet clouds \citep{2014ApJ...794...91D, 2016ApJ...832...91D}. The collision time scale in Figure \ref{fig:S1} (Section \ref{sec:results1}) is comparable to the time between the heat-up of a precursor body during the collision and the separation of single droplets in the collisional aftermath. As shown in Figure \ref{fig:S1}, this time scale is orders of magnitudes shorter than the local diffusion time scale of silicate material heated to high melt fractions, i.e., to the peak temperatures of chondrules. Therefore, the cooling droplets in the collisional aftermath preserved heterogeneous primordial nucleosynthetic signatures. Such distinct signatures are reported within chondrules of the same meteorite \citep{olsen2016magnesium,2016LPICo1921.6503B}. As a further contribution, these variable chondrule signatures can also originate from different impact events, which were then mixed together during reaccretion in a subsequent accretion-collision cycle. Repeated thermal recycling of chondrules is also in line with recent studies of microchondrule formation \citep{2016M&PS...51..235B}.

In summary, we would like to emphasize that our results regarding the metal-silicate separation, chemical equilibration and the generation of variations among chondrules generally apply to both type I (FeO-poor) and type II (FeO-rich) chondrules. The presence of Fe,Ni metal varies between different chondrite groups \citep{2014M&PS...49.2133D,2015M&PS...50...15S} and the total amount of Fe,Ni metal within and in the vicinity of chondrules may be related to the oxygen and sulfur fugacity of the precursor body and the surrounding gaseous medium. Importantly, \emph{any} Fe,Ni, FeS or chemical and isotopic heterogeneity present in precursor bodies before the chondrule-forming impact event would have been erased in planetesimals that experienced a magma ocean stage. As we have shown, however, it was possible to preserve these anomalies in (at maximum) partially molten precursor bodies that accreted from diverse nebular material.

\begin{figure*}[tbh]
    \centering
    \includegraphics[width=0.79\textwidth]{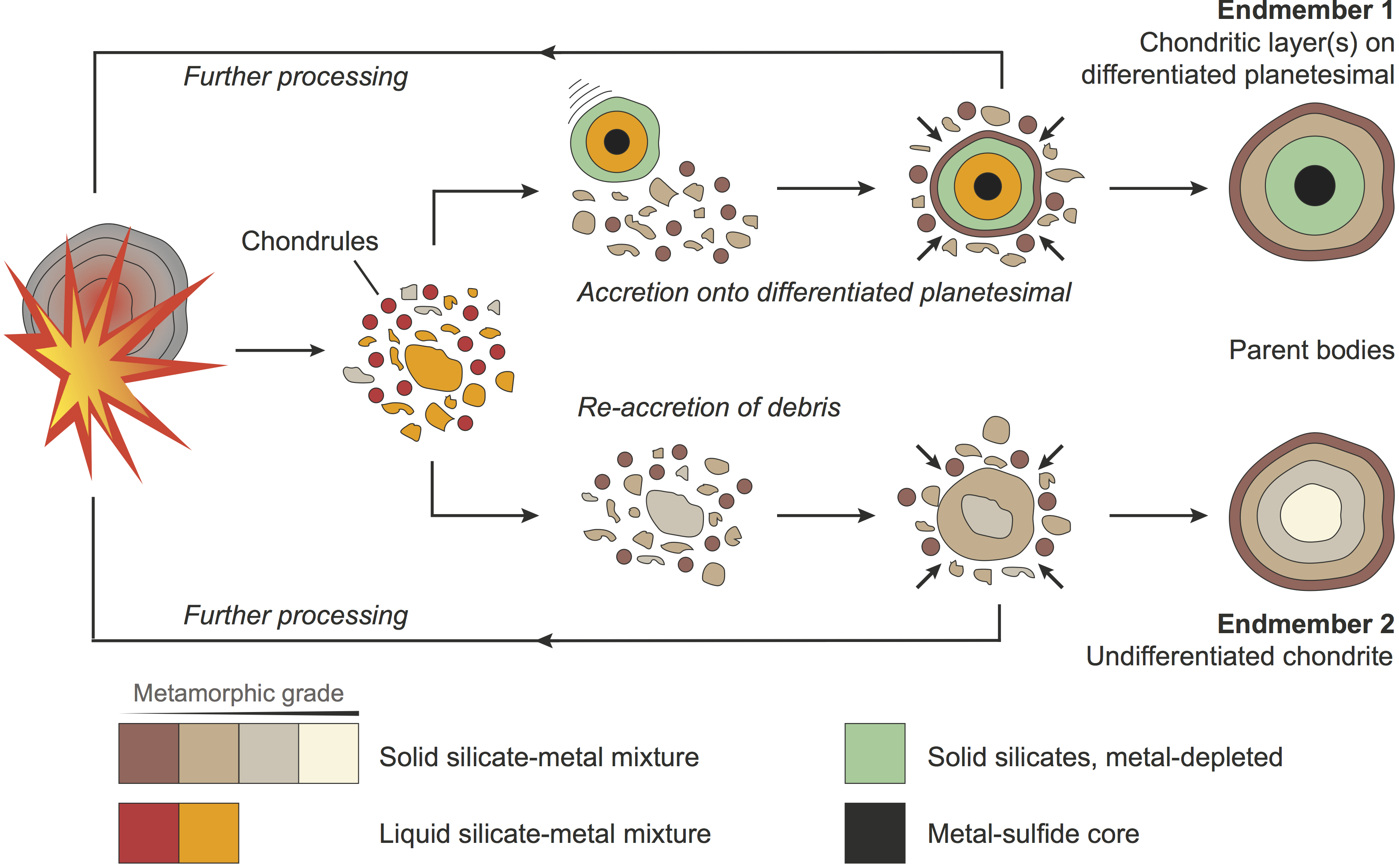}
    \caption{Schematic illustration of the accretion-collision cycles during which chondrules may form from impact splashes among radiogenically prehated planetesimals. The impacts launched expanding clouds of magma droplets \citep{2014ApJ...794...91D, 2016ApJ...832...91D} in addition to melted and unmelted debris, which subsequently cooled. The debris then either re-accumulated or accreted onto a neighboring planetesimal. Before forming the final chondrite parent bodies the material could go through multiple cycles of liberation and re-accumulation with varying degrees of injected energy and accumulation time scales. Cold matrix material (not shown) from the surrounding disk environment is accreted together with chondrules into the final chondrite parent body.}
    \label{fig:illustration2}
\end{figure*}

\subsection{Further constraints and outlook}
\label{sec:further}

Throughout this work we focused on geochemical and physical conditions for planetary materials {\em necessary} to retain abundant Fe,Ni metals, primordial isotopic and nucleosynthetic heterogeneities on a chondrule-size scale and to achieve the required peak temperatures for chondrule formation during planetesimal collisions. Further detailed work -- both from modelers and experimentalists -- is needed to investigate the enigmatic nature of chondrules and its link to the environment in the early solar nebula. For instance, under which circumstances can the post-collision droplet clouds satisfy the thermal histories and moderately volatile element retention  (e.g., Na and K) of chondrules \citep{2008Sci...320.1617A, 2014ApJ...794...91D, 2016ApJ...832...91D}, in case the thermal histories derived so far are reliable \citep{2017LPICo1963.2008L}? Other important issues relate to, e.g., the prevalence of porphyritc textures among chondrules, varying chondrule distributions among different chondrite groups or the retention of relict grains. So far, there are only a few examples of chondrules which show strong experimental evidence for being generated by an impact  \citep{2005Natur.436..989K,2016SciA....2E1001M}. More detailed work on the thermo-physical conditions during and after planetesimal impacts needs to be undertaken to compare theoretical expectations with experimental evidence with the goal of a {\em sufficient} set of evidence to either strengthen or rule out impacts as a formation mechanism for the majority of chondrules.

%%%%%%%%%%%%%%%%%%%%%%%%%%%%%%%%%%%%%%%%%%%%%%%%%%%%%%%%%%%%%%%%%%%%%%%%%%%%%%%%%%%%%%%%%%%%%%
%%%%%%%%%%%%%%%%%%%%%%%%%%%%%%%%%%%%%%%%%%%%%%%%%%%%%%%%%%%%%%%%%%%%%%%%%%%%%%%%%%%%%%%%%%%%%%
%%%%%%%%%%%%%%%%%%%%%%%%%%%%%%%%%%%%%%%%%%%%%%%%%%%%%%%%%%%%%%%%%%%%%%%%%%%%%%%%%%%%%%%%%%%%%%
%%%%%%%%%%%%%%%%%%%%%%%%%%%%%%%%%%%%%%%%%%%%%%%%%%%%%%%%%%%%%%%%%%%%%%%%%%%%%%%%%%%%%%%%%%%%%%
%%%%%%%%%%%%%%%%%%%%%%%%%%%%%%%%%%%%%%%%%%%%%%%%%%%%%%%%%%%%%%%%%%%%%%%%%%%%%%%%%%%%%%%%%%%%%%
%%%%%%%%%%%%%%%%%%%%%%%%%%%%%%%%%%%%%% CONCLUSIONS %%%%%%%%%%%%%%%%%%%%%%%%%%%%%%%%%%%%%%%%%%%
%%%%%%%%%%%%%%%%%%%%%%%%%%%%%%%%%%%%%%%%%%%%%%%%%%%%%%%%%%%%%%%%%%%%%%%%%%%%%%%%%%%%%%%%%%%%%%
%%%%%%%%%%%%%%%%%%%%%%%%%%%%%%%%%%%%%%%%%%%%%%%%%%%%%%%%%%%%%%%%%%%%%%%%%%%%%%%%%%%%%%%%%%%%%%
%%%%%%%%%%%%%%%%%%%%%%%%%%%%%%%%%%%%%%%%%%%%%%%%%%%%%%%%%%%%%%%%%%%%%%%%%%%%%%%%%%%%%%%%%%%%%%
%%%%%%%%%%%%%%%%%%%%%%%%%%%%%%%%%%%%%%%%%%%%%%%%%%%%%%%%%%%%%%%%%%%%%%%%%%%%%%%%%%%%%%%%%%%%%%
%%%%%%%%%%%%%%%%%%%%%%%%%%%%%%%%%%%%%%%%%%%%%%%%%%%%%%%%%%%%%%%%%%%%%%%%%%%%%%%%%%%%%%%%%%%%%%
\section{Conclusions}
\label{sec:conclusions}

In this manuscript we examined the formation of chondrules from collisions of planetesimals, which were preheated from the radioactive decay of \al. First, we investigated the end-member scenario of collisions between planetesimals heated to silicate melt fractions above the rheological transition, i.e., with interior magma oceans. Using well-studied scaling relations of the metal rainfall mechanism and the local diffusion time scale in convective silicate systems we determined that such planetesimals
\begin{enumerate}[(i)]
\item cannot suspend significant amounts of Fe,Ni metal and, therefore, evolve to a physically 
differentiated structure
\item[] and
\item rapidly equilibrated primordial chemical and nucleosynthetic heterogeneities.
\end{enumerate}

Therefore, we conclude that physically plausible impact splash interactions between such bodies would have resulted in chondrule-like but basaltic spherules, which are not observed in the meteoritic record. Contrary to \citet{asphaug2017}, we argue that this is a telltale-sign that no such interactions took place in the early solar system and that planetesimals with large-scale interior magma oceans were not abundant in the source reservoir of today's asteroid main belt. Potential reasons for this may be delayed planetesimal formation, sub-canonical \al abundances in the planetesimal formation region, efficient heat source redistribution by migration of aluminum-rich melts to the surface \citep{wilson_keil_2017}, or widespread collisional interactions of fully-molten bodies were prevented by environmental (disk conditions) or dynamical (collisional growth-related) mechanisms, for instance, efficient collisional recycling in the source reservoir of nowadays asteroid belt. Furthermore, it would imply that iron meteorites would have been primarily formed via incomplete metal-silicate differentiation or must originate from larger bodies than currently anticipated \citep{2017LPI....48.2433L}. We suggest that these conclusions can help to achieve a better understanding of the early dynamical environment during the solar protoplanetary disk phase, because it excludes the part of the parameter space that leads to widespread generation of droplet-like and basaltic material feeding the asteroid main belt.

We argue that the debate of a collisional (or `planetary') origin of chondrules needs to take into account the complications of the combined planetesimal evolution and recycling efficiency during accretion. The early formation and reaccretion of planetesimals of low mass and/or under sub-canonical \al abundances opens the window to a vast collisional parameter space, which may satisfy many geo- and cosmochemical constraints derived from the meteoritic record. We have sketched one such accretion-collision cycle to generate chondrules in Figure \ref{fig:illustration2}.  In the future, the dynamical feasibility and implications of our proposed chondrule formation scenario can be explored with astrophysical models that simultaneously solve for a global planetesimal source system and achieve sufficiently high mass resolution to resolve the low-mass bodies we focused on in this work \citep[e.g.,][]{2012AJ....144..119L,2017Icar281459M}.

In summary, we propose that the linkage of the initial planetesimal size-frequency distribution, formation time, interior evolution and collisional recycling may be further used to constrain the formation of chondrules and subsequently the chondrite parent bodies. The collisional chondrule formation scenario links the chondrule origin to the formation of the terrestrial planets and the solar system architecture we observe today. Details of the model -- such as the exact disk conditions necessary to create such an environment and the thermo-physical processes and energy localization during collisions -- demand detailed physical and chemical models on many spatial and temporal scales, which offer exciting new pathways for the study of planet formation. These models need to be further synchronized and tested against precise laboratory data and may ultimately lead the way to a better understanding of the earliest environment of the solar nebula.

\vspace{5mm}
\paragraph{Acknowledgements} T.L. gratefully acknowledges insightful discussions with H. Palme, J. Zipfel, N. P. Walte, J. Dr\c{a}\.{z}kowska, S. N. Raymond and artistic advice from A. Rozel and C. Jain. The authors thank Z. M. Leinhardt and P. J. Carter for providing us a \textsc{python}-based script to evaluate planetesimal interactions on the basis of the \textsc{edacm} scaling relations \citep{2012ApJ74579L}, C. M. O'D. Alexander and S. J. Desch for thorough and constructive reviews that considerably helped to improve the manuscript, A. Morbidelli for the editorial handling, and the organisers of the 2015 Gordon Research Conference `Origins of Solar Systems' for an inspiring and collaborative meeting that initiated this project. The models were analyzed using the open source software environment \textsc{matplotlib} \citep{matplotlib}. T.L. was supported by ETH Research Grant ETH-17 13-1. The numerical simulations in this work were performed on the \textsc{euler} computing cluster of ETH Z{\"u}rich. Parts of this work have been carried out within the framework of the National Center for Competence in Research PlanetS supported by the Swiss National Science Foundation.
\vspace{5mm}

% \section*{Bibliography}

% \bibliographystyle{elsarticle-harv} 
% \bibliography{bibliography} % 

%%%%%%%%%%%%%%%%%%%%%%%%%%%%%%%%%%%%%%%%%%%%%%%%%%%%%%%%%%%%%%%%%%%%%%%%%%%%%%%%%%%%%%%%%%%%%%
%%%%%%%%%%%%%%%%%%%%%%%%%%%%%%%%%%%%%%%%%%%%%%%%%%%%%%%%%%%%%%%%%%%%%%%%%%%%%%%%%%%%%%%%%%%%%%
%%%%%%%%%%%%%%%%%%%%%%%%%%%%%%%%%%%%%%%%%%%%%%%%%%%%%%%%%%%%%%%%%%%%%%%%%%%%%%%%%%%%%%%%%%%%%%
%%%%%%%%%%%%%%%%%%%%%%%%%%%%%%%%%%%%%%%%%%%%%%%%%%%%%%%%%%%%%%%%%%%%%%%%%%%%%%%%%%%%%%%%%%%%%%
%%%%%%%%%%%%%%%%%%%%%%%%%%%%%%%%%%%%%%%%%%%%%%%%%%%%%%%%%%%%%%%%%%%%%%%%%%%%%%%%%%%%%%%%%%%%%%
%%%%%%%%%%%%%%%%%%%%%%%%%%%%%%%%%%%%%%% APPENDIX %%%%%%%%%%%%%%%%%%%%%%%%%%%%%%%%%%%%%%%%%%%%%
%%%%%%%%%%%%%%%%%%%%%%%%%%%%%%%%%%%%%%%%%%%%%%%%%%%%%%%%%%%%%%%%%%%%%%%%%%%%%%%%%%%%%%%%%%%%%%
%%%%%%%%%%%%%%%%%%%%%%%%%%%%%%%%%%%%%%%%%%%%%%%%%%%%%%%%%%%%%%%%%%%%%%%%%%%%%%%%%%%%%%%%%%%%%%
%%%%%%%%%%%%%%%%%%%%%%%%%%%%%%%%%%%%%%%%%%%%%%%%%%%%%%%%%%%%%%%%%%%%%%%%%%%%%%%%%%%%%%%%%%%%%%
%%%%%%%%%%%%%%%%%%%%%%%%%%%%%%%%%%%%%%%%%%%%%%%%%%%%%%%%%%%%%%%%%%%%%%%%%%%%%%%%%%%%%%%%%%%%%%
%%%%%%%%%%%%%%%%%%%%%%%%%%%%%%%%%%%%%%%%%%%%%%%%%%%%%%%%%%%%%%%%%%%%%%%%%%%%%%%%%%%%%%%%%%%%%%
\appendix

\onecolumn

% http://www.stat.berkeley.edu/~paciorek/computingTips/Customizing_numbering_pages.html
% \renewcommand{\thepage}{S\arabic{page}}  
% \renewcommand{\thesection}{S\arabic{section}}   
% \renewcommand{\thetable}{S\arabic{table}}   
% \renewcommand{\thefigure}{S\arabic{figure}}
% \renewcommand{\theequation}{S\arabic{equation}}
% \renewcommand{\figurename}{Supplemental Material, Figure}
\setcounter{figure}{0}  % reset counter

\section{Supplementary Figures}
\label{sec:suppl}

\begin{figure}[htb]
    \centering
    \includegraphics[width=0.95\textwidth]{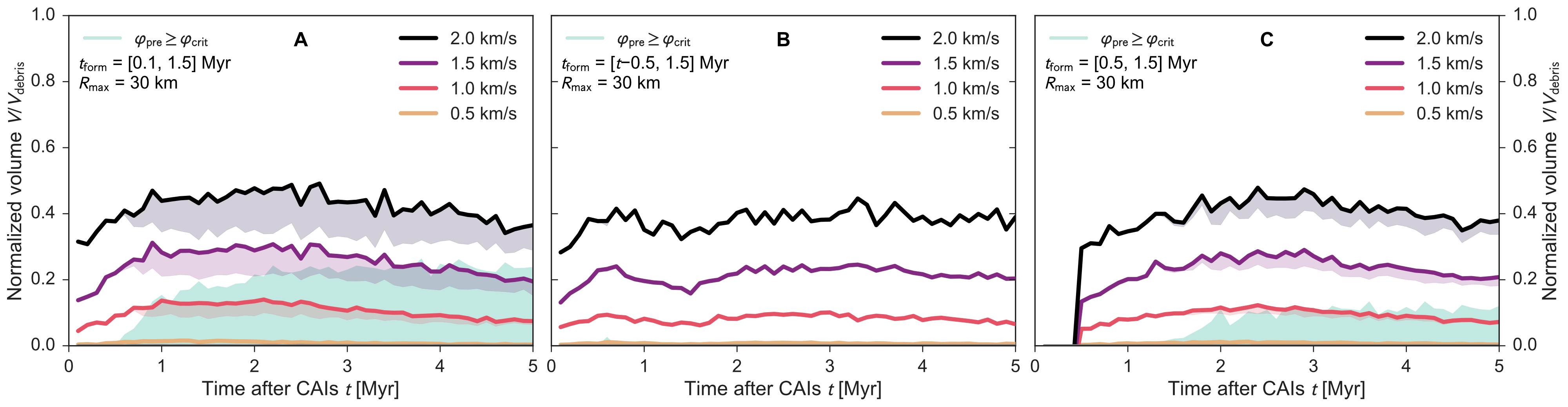}
    \caption{Simulation outputs for planetesimal families with radii \Rmax = 30 km and formation time regimes \tform = {\bf (A)} [0.1, 1.5], {\bf (B)} [t-0.5, 1.5] and {\bf (C)} [0.5, 1.5] Myr. See Figure \ref{fig:collision1} for a detailed description.}
    \label{fig:S3}
\end{figure}

\begin{figure}[htb]
    \centering
    \includegraphics[width=0.95\textwidth]{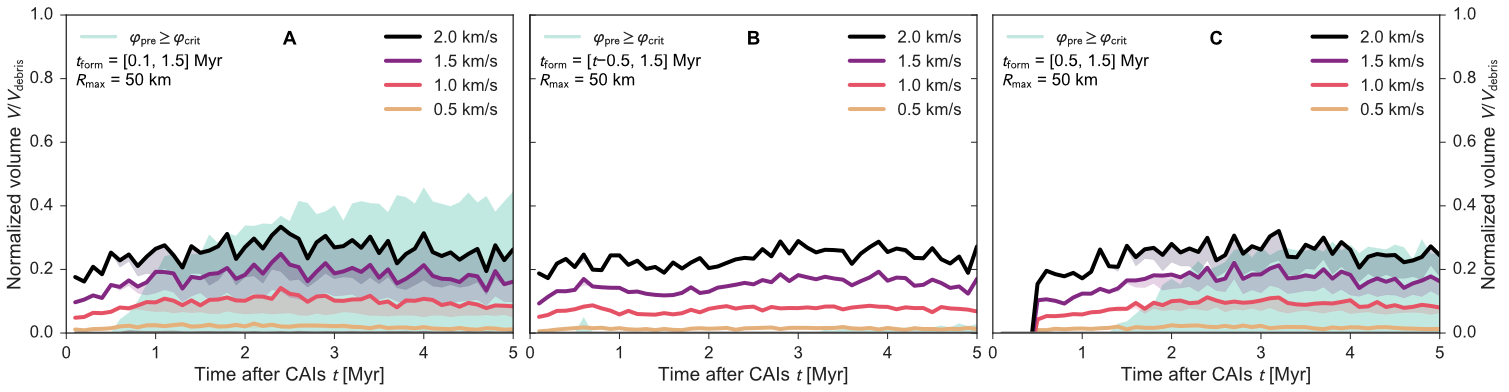}
    \caption{Simulation outputs for planetesimal families with radii \Rmax = 50 km and formation time regimes as in Figure \ref{fig:S3}.}
    \label{fig:S4}
\end{figure} 

\begin{figure}[htb]
    \centering
    \includegraphics[width=0.95\textwidth]{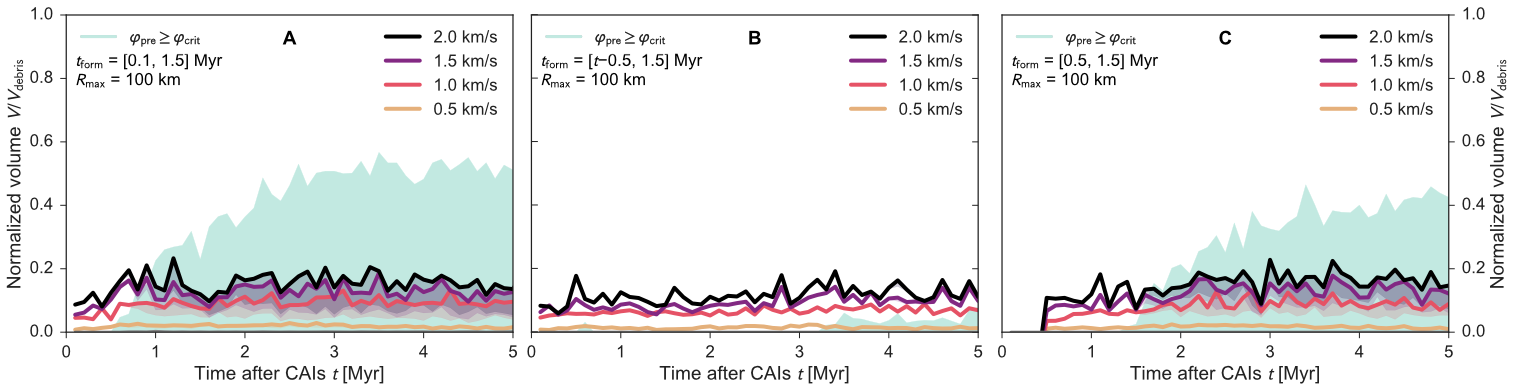}
    \caption{Simulation outputs for planetesimal families with radii \Rmax = 100 km and formation time regimes as in Figure \ref{fig:S3}.}
    \label{fig:S5}
\end{figure} 

\begin{figure}[htb]
    \centering
    \includegraphics[width=0.33\textwidth]{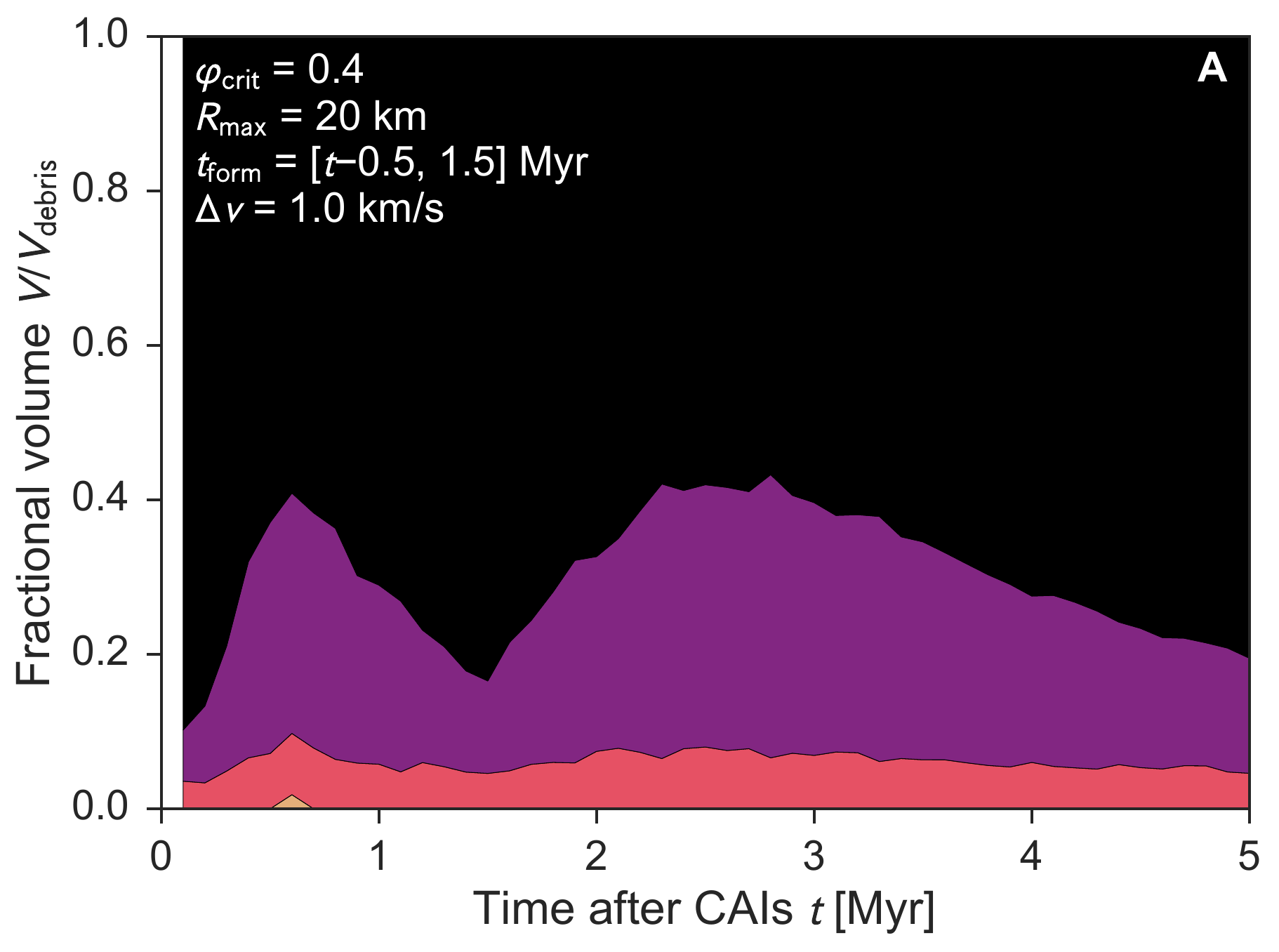}
    \includegraphics[width=0.33\textwidth]{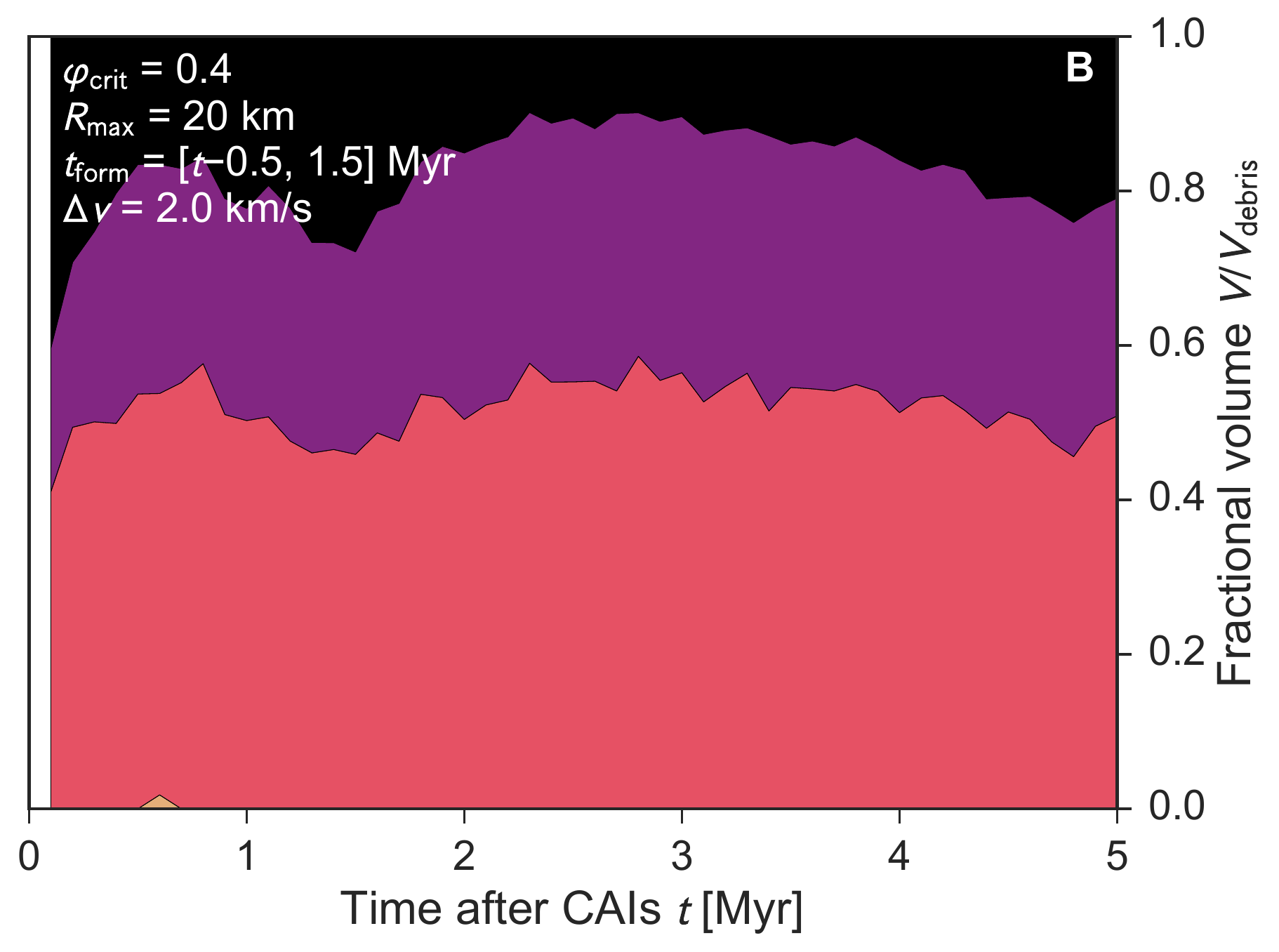}
    \caption{Thermal debris distribution for collision models with \Rmax = 20 km, \tform = [t-0.5, 1.5] Myr and collision velocities of {\bf(A)} 1.0 km/s and {\bf(B)} 2.0 km/s. Black represents unmelted, purple partially melted, red chondrule-eligible (\Tpost $>$ \Tchondrule) and yellow metal-depleted material. (Compare Figure \ref{fig:thermal1}.)}
    \label{fig:S6}
\end{figure}

\begin{figure}[htb]
    \centering
    \includegraphics[width=0.33\textwidth]{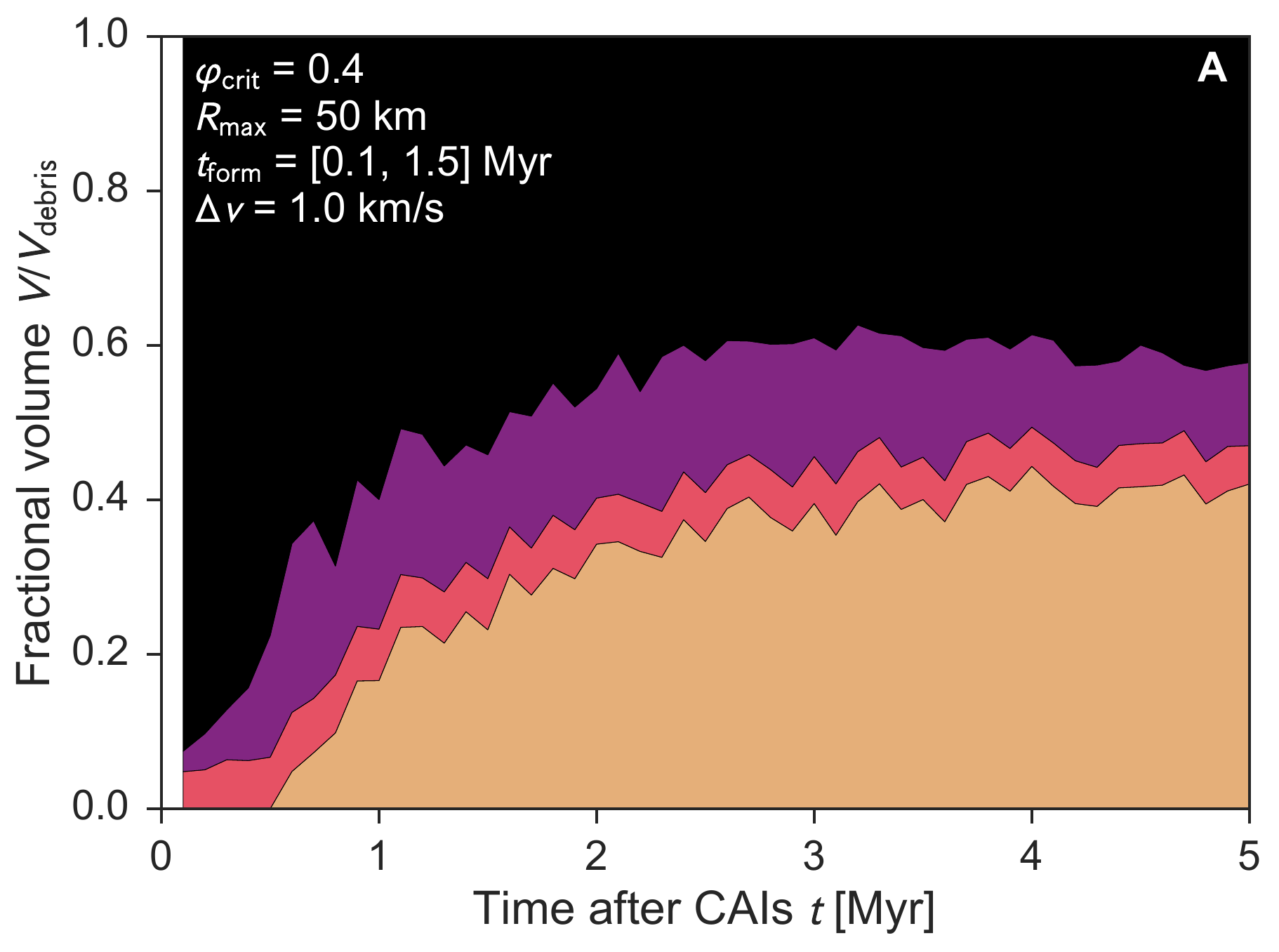}
    \includegraphics[width=0.33\textwidth]{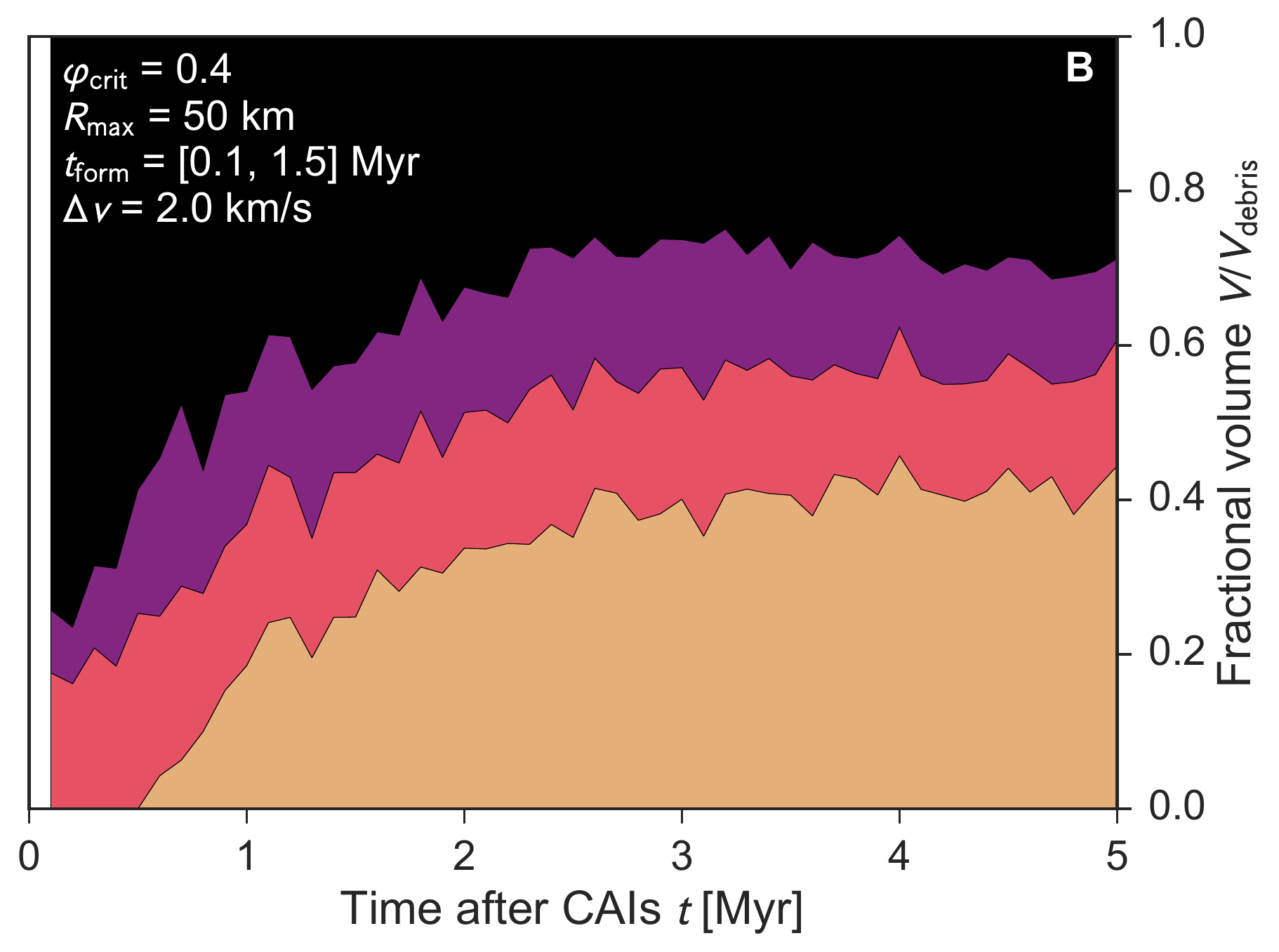}
    \caption{Thermal debris distribution for collision models with \Rmax = 50 km, \tform = [0.1, 1.5] Myr and collision velocities of {\bf(A)} 1.0 km/s and {\bf(B)} 2.0 km/s. Black represents unmelted, purple partially melted, red chondrule-eligible (\Tpost $>$ \Tchondrule) and yellow metal-depleted material. (Compare Figure \ref{fig:thermal1}.)}
    \label{fig:S8}
\end{figure}
 
\begin{figure}[htb]
    \centering
    \includegraphics[width=0.33\textwidth]{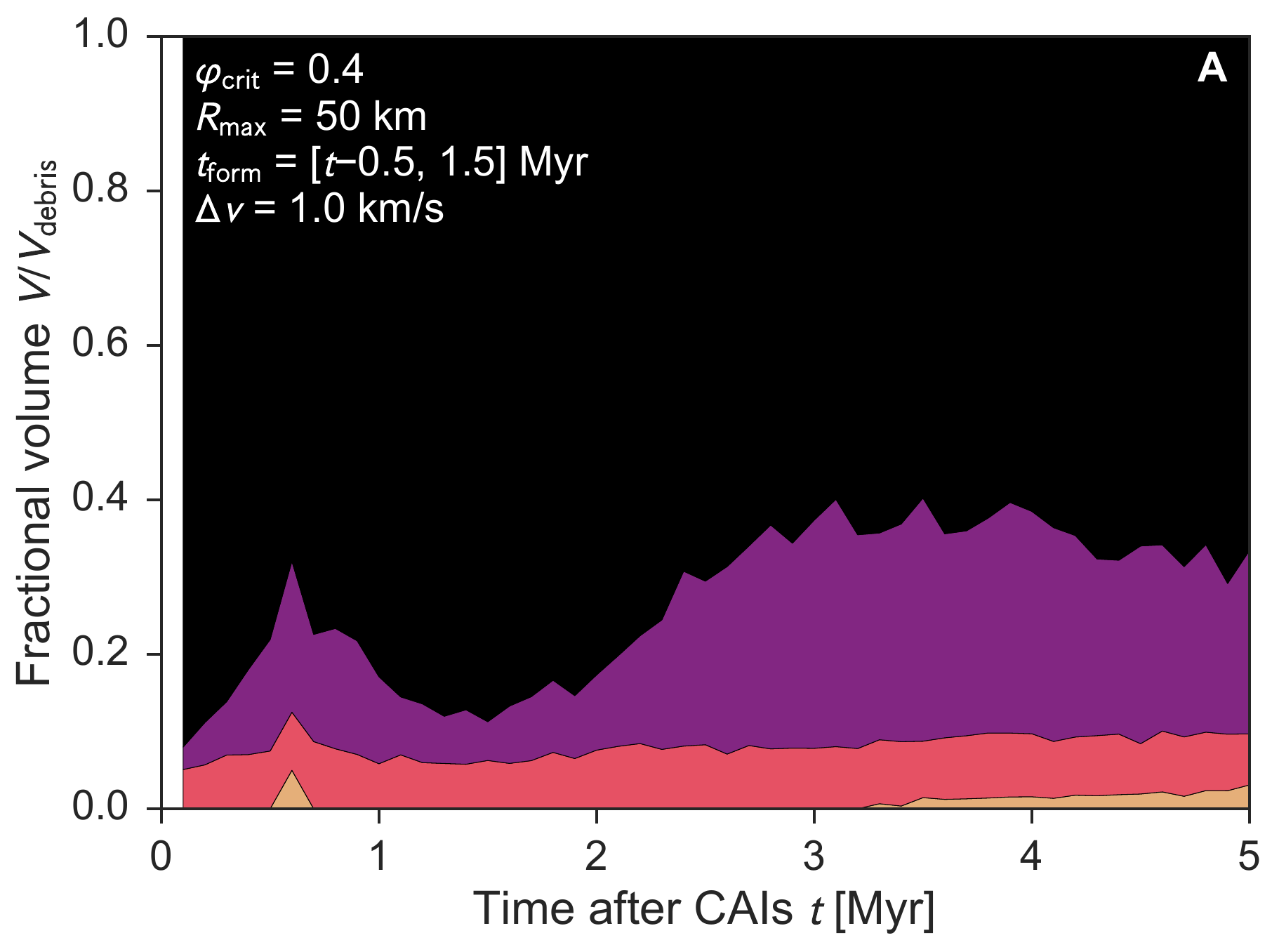}
    \includegraphics[width=0.33\textwidth]{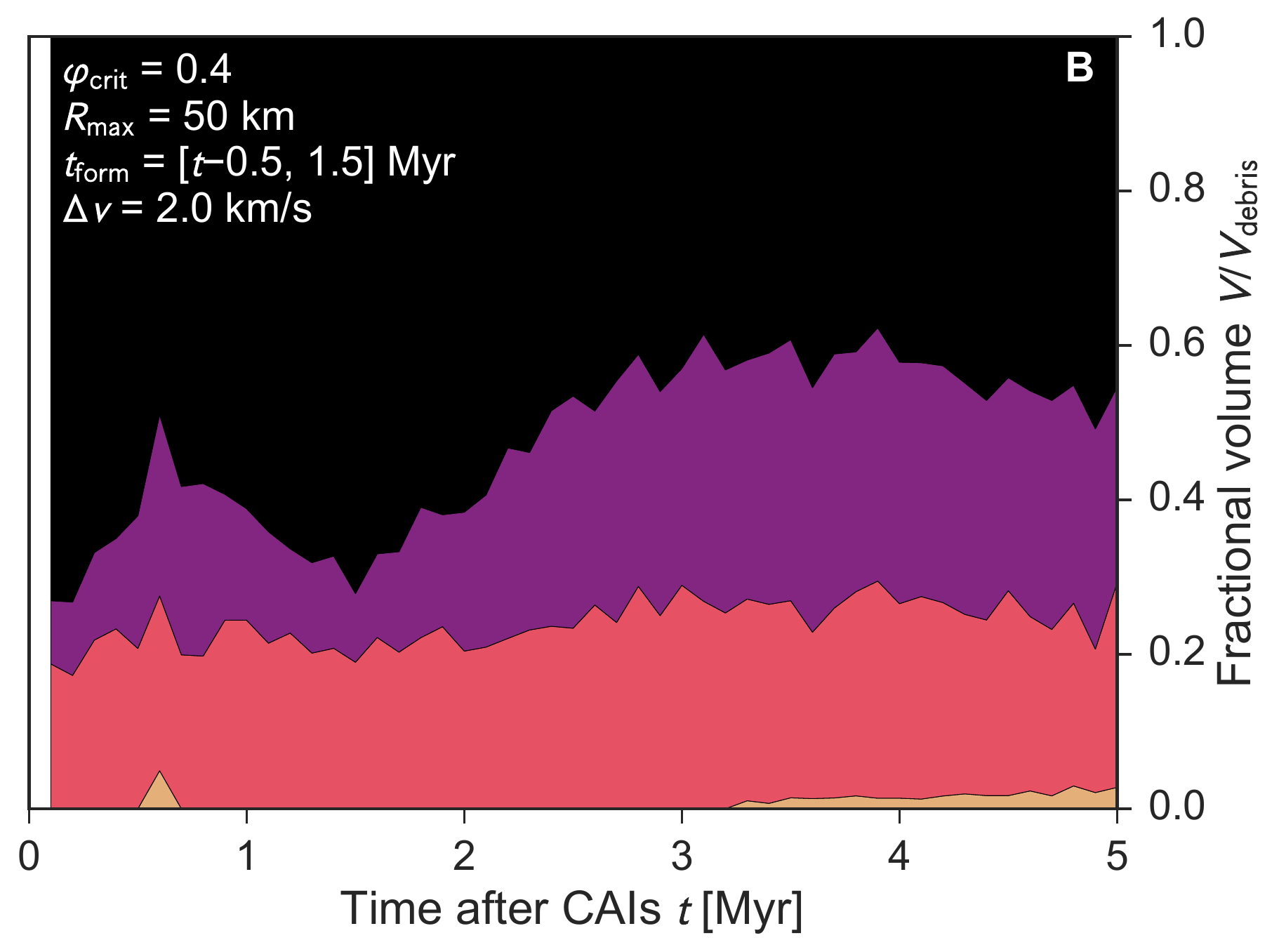}
    \caption{Thermal debris distribution for collision models with \Rmax = 50 km, \tform = [t-0.5, 1.5] Myr and collision velocities of {\bf(A)} 1.0 km/s and {\bf(B)} 2.0 km/s. Black represents unmelted, purple partially melted, red chondrule-eligible (\Tpost $>$ \Tchondrule) and yellow metal-depleted material. (Compare Figure \ref{fig:thermal1}.)}
    \label{fig:S9}
\end{figure}

\begin{figure}[htb]
    \centering
    \includegraphics[width=0.33\textwidth]{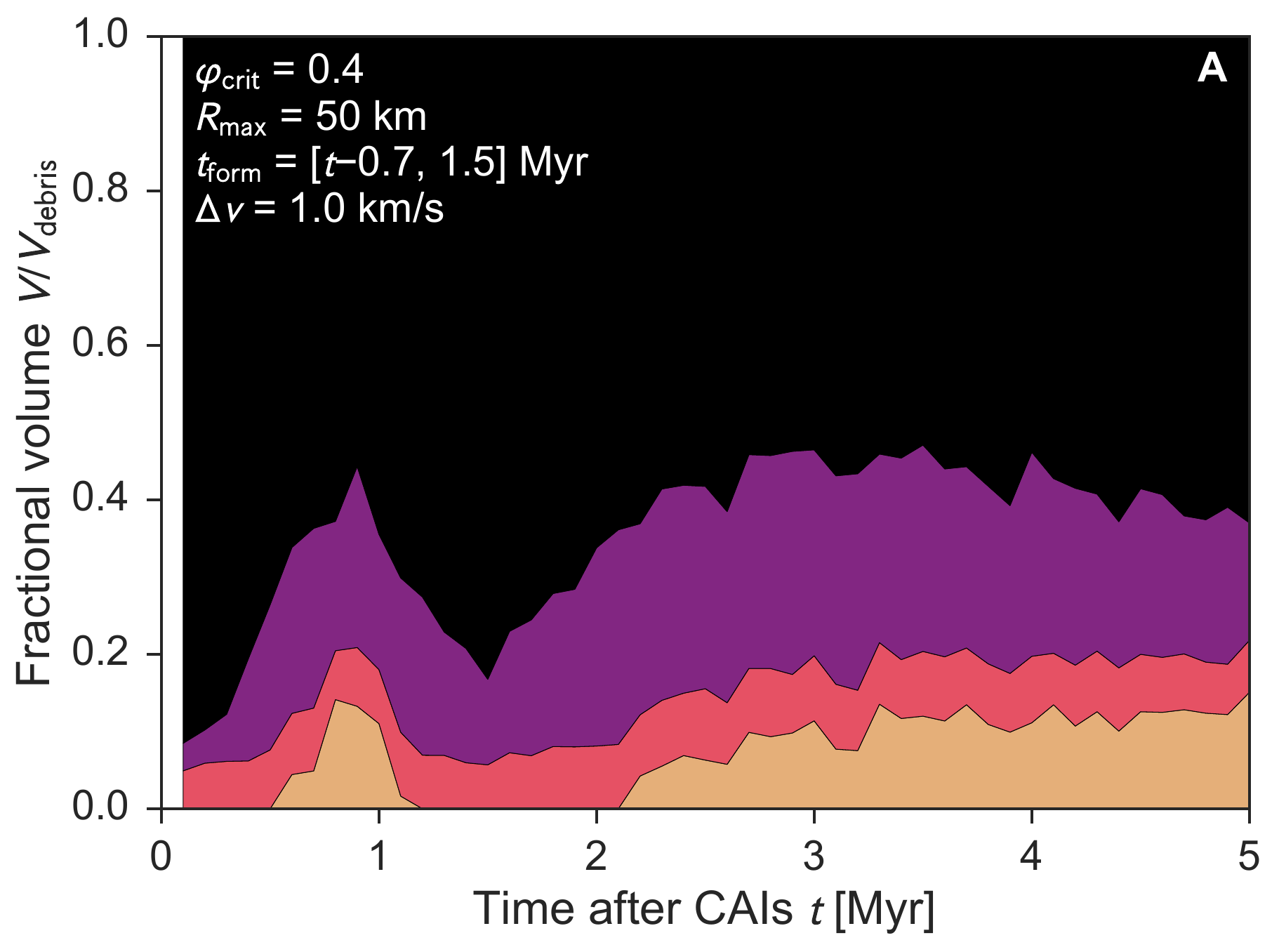}
    \includegraphics[width=0.33\textwidth]{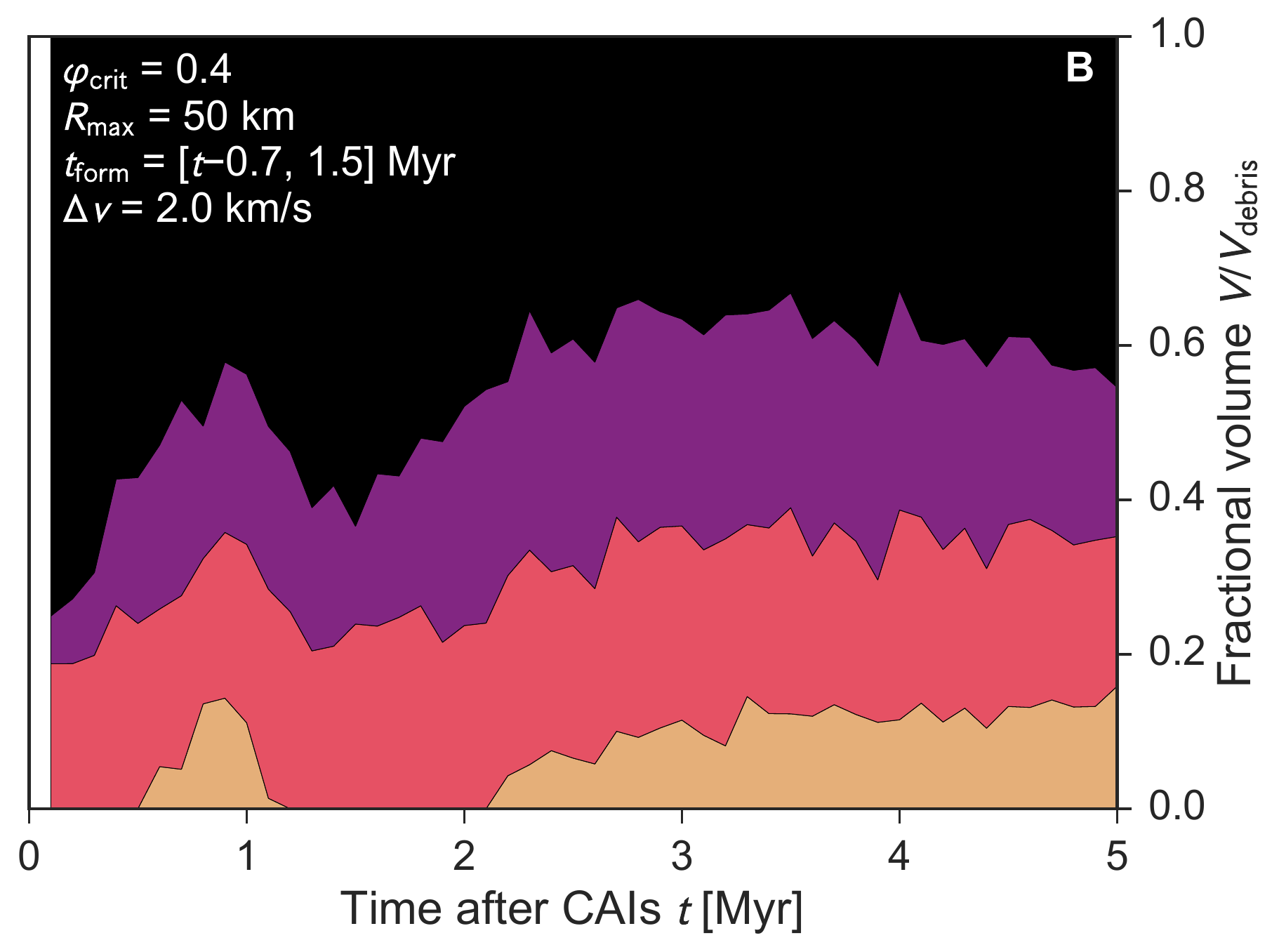}
    \caption{Thermal debris distribution for collision models with \Rmax = 50 km, \tform = [t-0.7, 1.5] Myr and collision velocities of {\bf(A)} 1.0 km/s and {\bf(B)} 2.0 km/s. Black represents unmelted, purple partially melted, red chondrule-eligible (\Tpost $>$ \Tchondrule) and yellow metal-depleted material. (Compare Figure \ref{fig:thermal1}.)}
    \label{fig:S10}
\end{figure}

\begin{figure}[htb]
    \centering
    \includegraphics[width=0.33\textwidth]{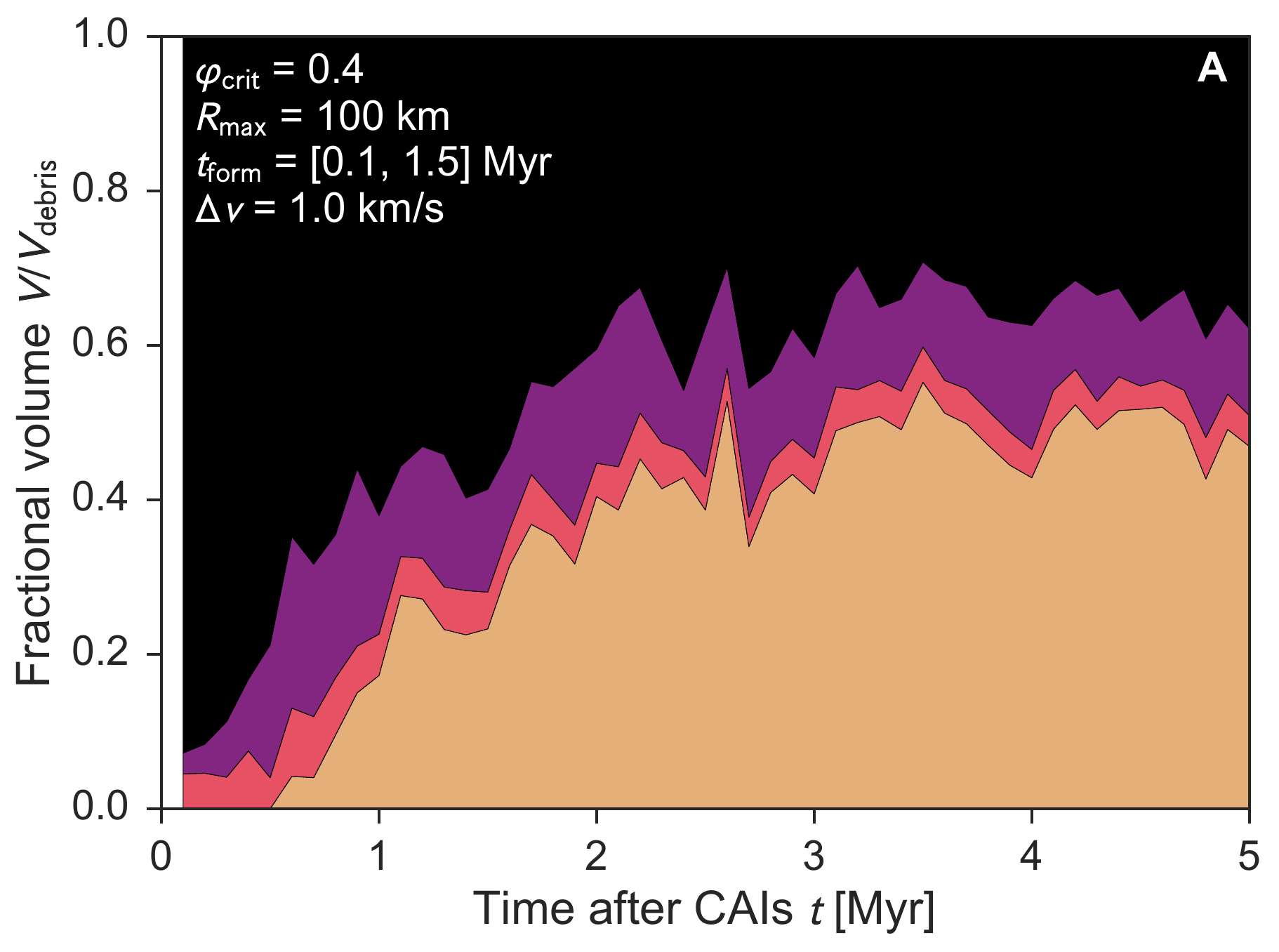}
    \includegraphics[width=0.33\textwidth]{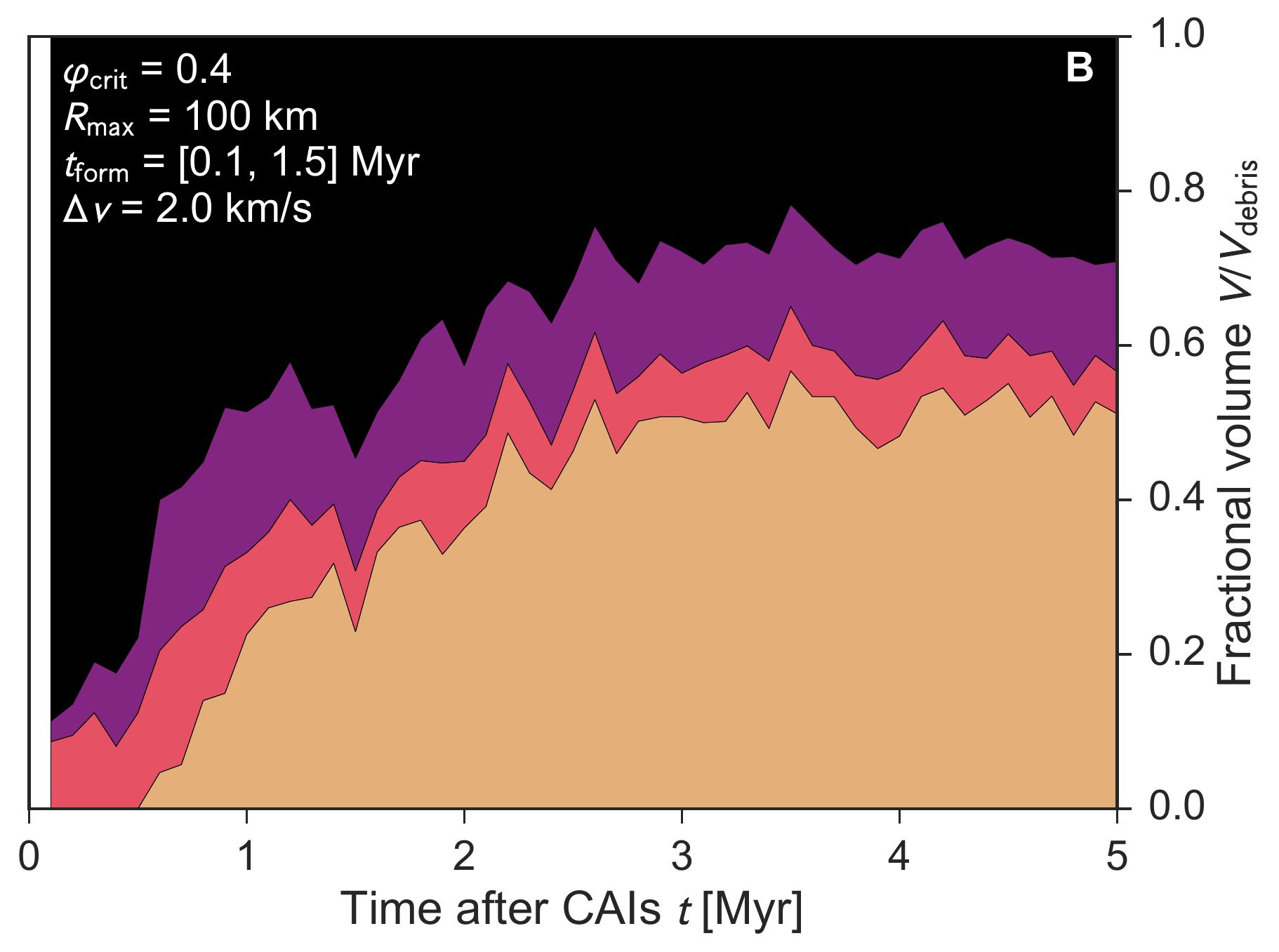}
    \caption{Thermal debris distribution for collision models with \Rmax = 100 km, \tform = [0.1, 1.5] Myr and collision velocities of {\bf(A)} 1.0 km/s and {\bf(B)} 2.0 km/s. Black represents unmelted, purple partially melted, red chondrule-eligible (\Tpost $>$ \Tchondrule) and yellow metal-depleted material. (Compare Figure \ref{fig:thermal1}.)}
    \label{fig:S11}
\end{figure}

\begin{figure}[htb]
    \centering
    \includegraphics[width=0.33\textwidth]{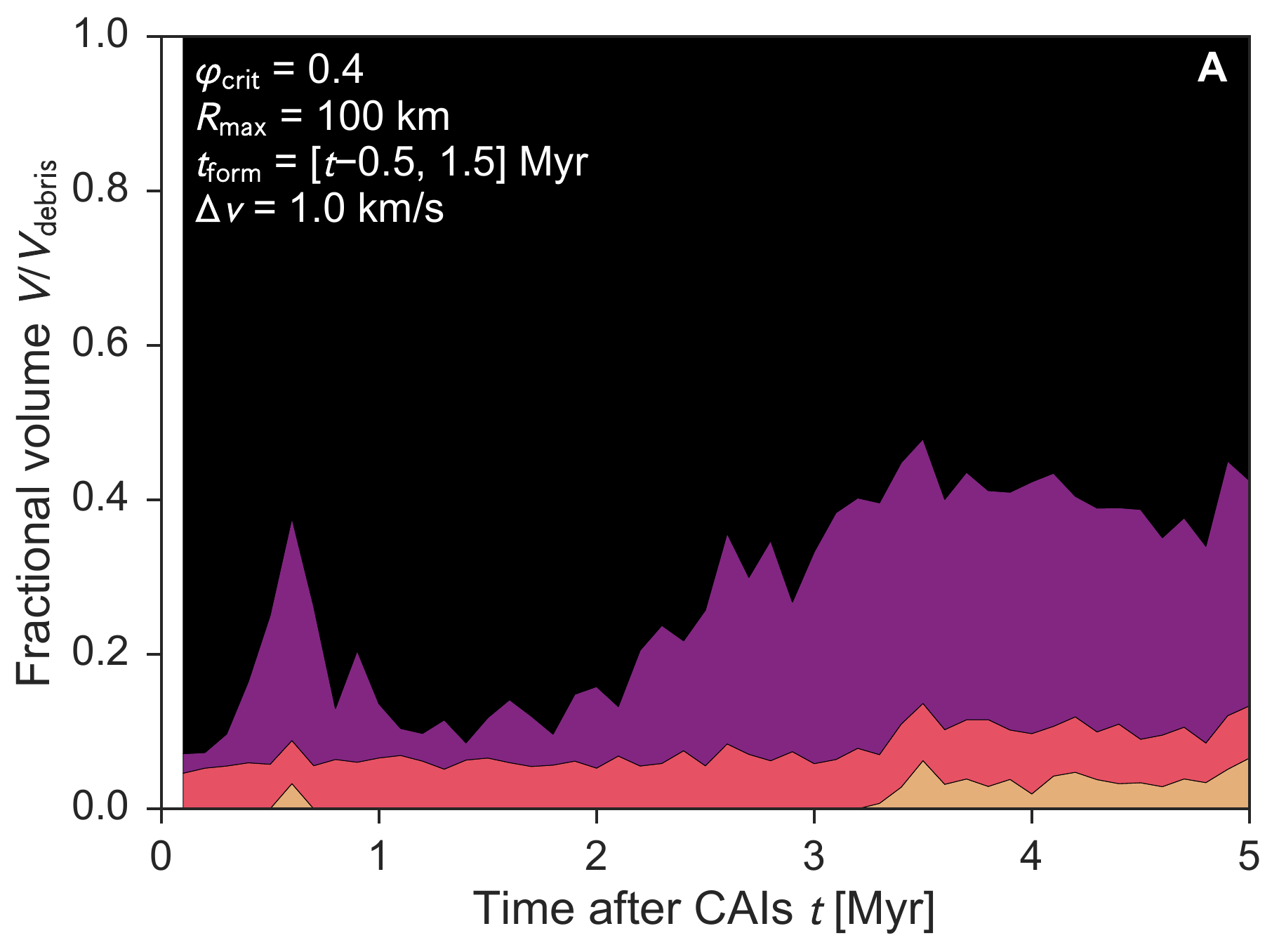}
    \includegraphics[width=0.33\textwidth]{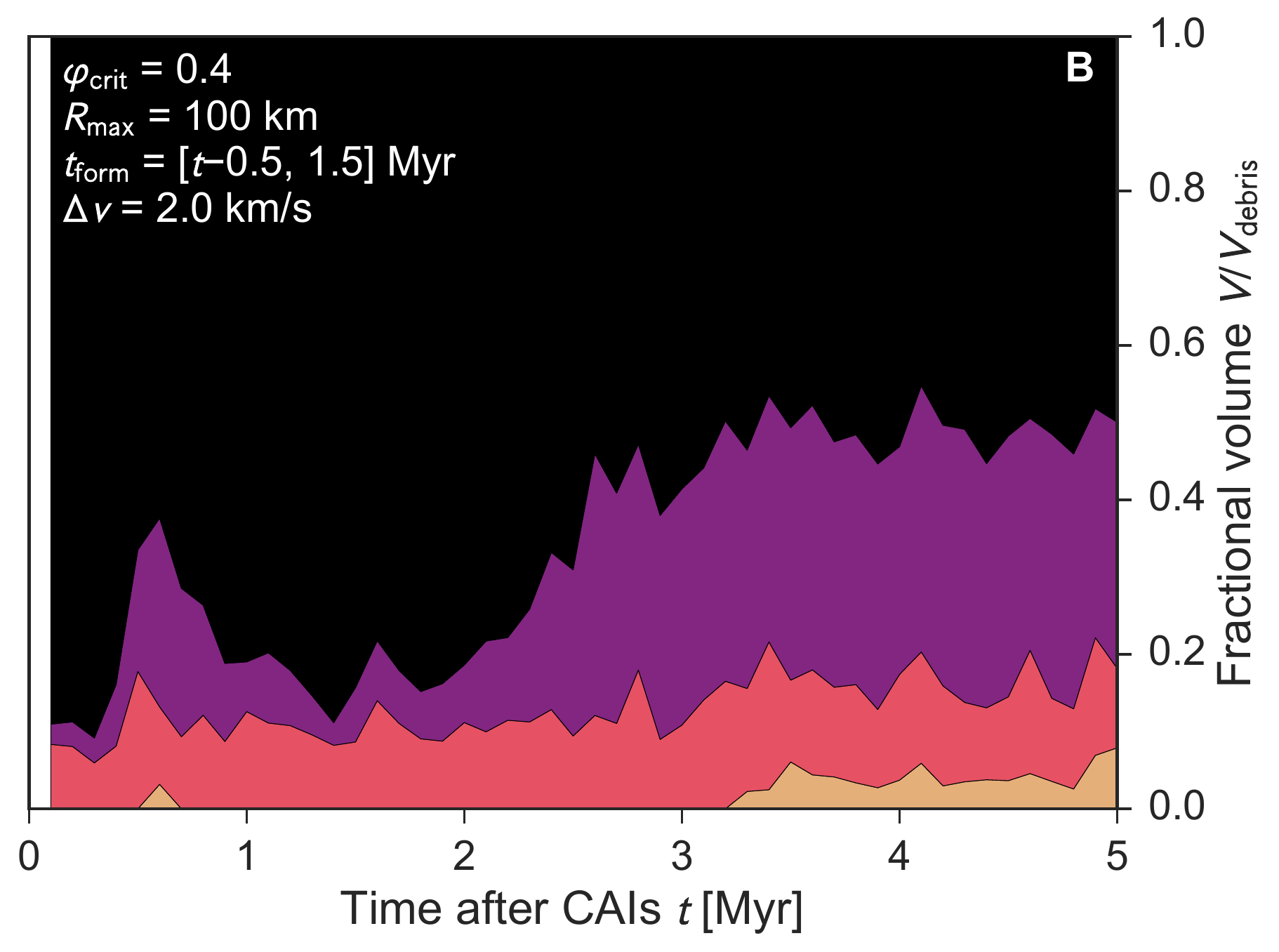}
    \caption{Thermal debris distribution for collision models with \Rmax = 100 km, \tform = [t-0.5, 1.5] Myr and collision velocities of {\bf(A)} 1.0 km/s and {\bf(B)} 2.0 km/s. Black represents unmelted, purple partially melted, red chondrule-eligible (\Tpost $>$ \Tchondrule) and yellow metal-depleted material. (Compare Figure \ref{fig:thermal1}.)}
    \label{fig:S12}
\end{figure}

\begin{figure}[htb]
    \centering
    \includegraphics[width=0.33\textwidth]{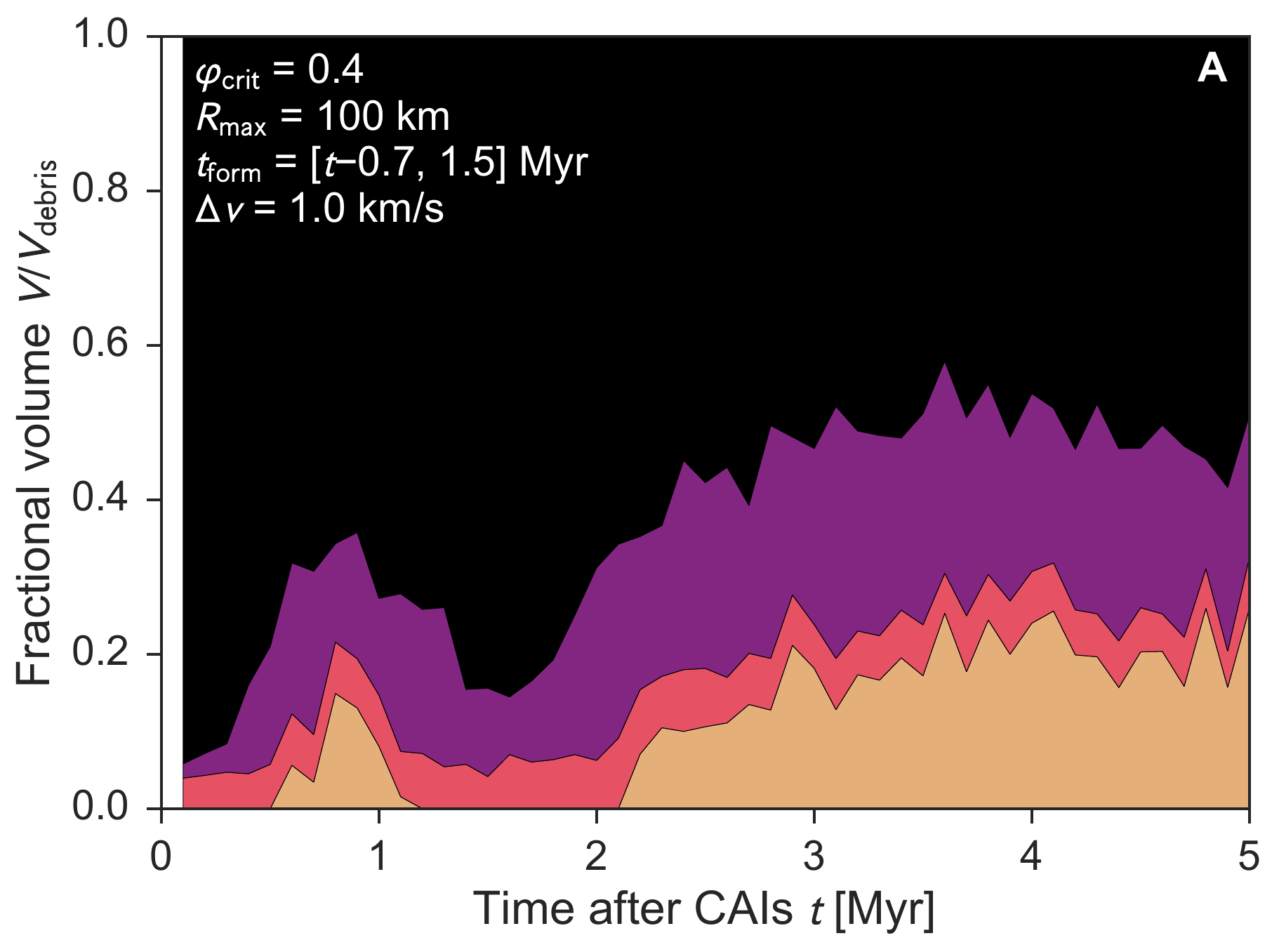}
    \includegraphics[width=0.33\textwidth]{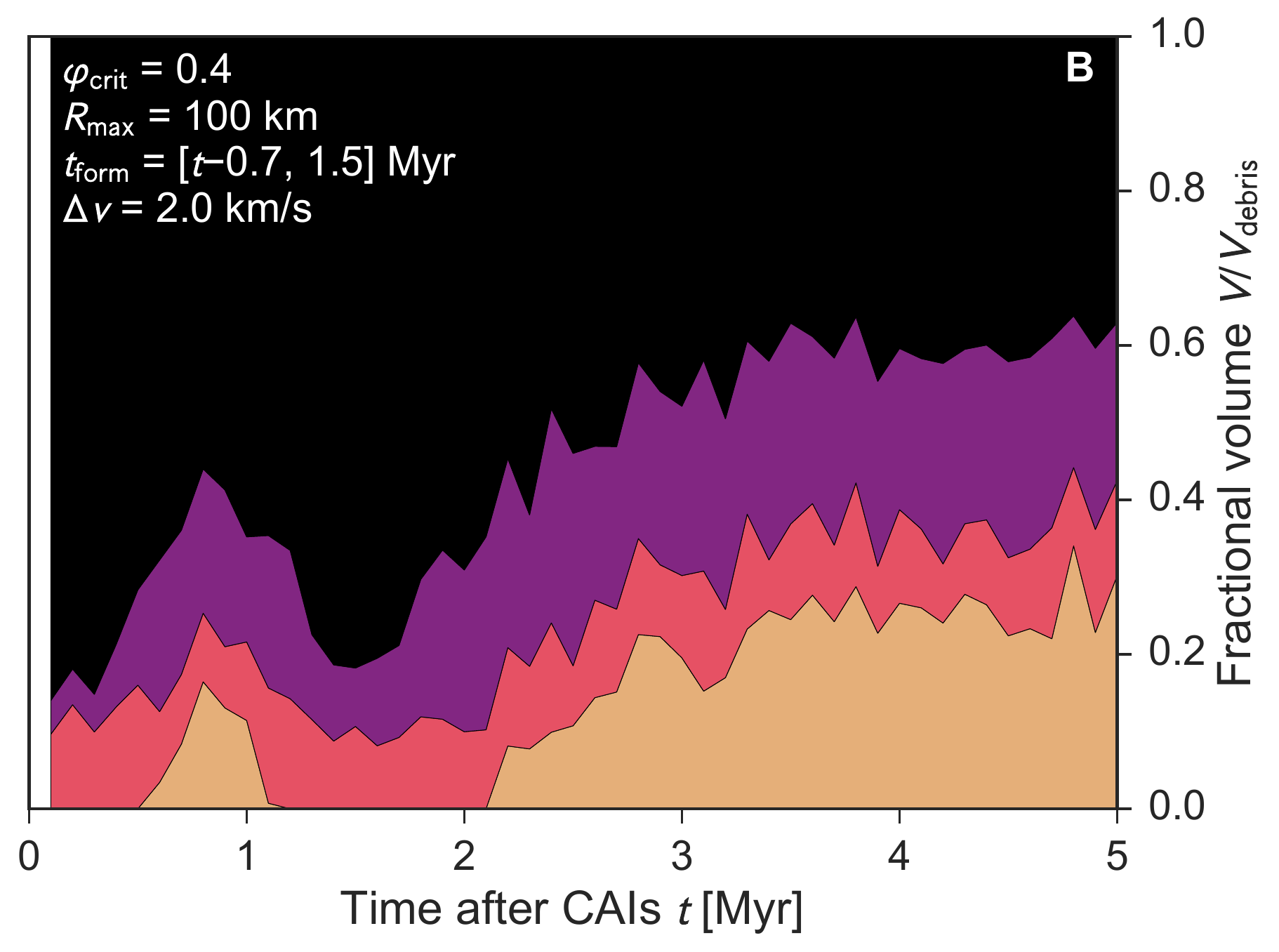}
    \caption{Thermal debris distribution for collision models with \Rmax = 100 km, \tform = [t-0.7, 1.5] Myr and collision velocities of {\bf(A)} 1.0 km/s and {\bf(B)} 2.0 km/s. Black represents unmelted, purple partially melted, red chondrule-eligible (\Tpost $>$ \Tchondrule) and yellow metal-depleted material. (Compare Figure \ref{fig:thermal1}.)}
    \label{fig:S13}
\end{figure}

\end{document}